\documentclass[journal]{IEEEtran}
\usepackage{cite}
\usepackage{graphicx}
\usepackage{pifont}
\usepackage{amsmath,amssymb,amsthm}
\usepackage[font=footnotesize,skip=5pt,compatibility=false]{caption}
\usepackage{relsize}
\usepackage{makecell}
\usepackage{mathtools}
\usepackage{algorithmic}
\usepackage{algorithm}
\usepackage{adjustbox}
\usepackage{epstopdf}
\usepackage{enumitem}
\usepackage{tabularx}
\usepackage{lipsum}
\usepackage{hyperref}
\usepackage{svg}
\usepackage[most]{tcolorbox}
\usepackage{float}        % For better float control
\usepackage{placeins}     % To use \FloatBarrier
\usepackage{multicol}
\usepackage{graphicx,float}
\usepackage{xurl}
\usepackage{hyperref}
\usepackage{pgfplots}
\usepackage{subcaption}
\usepackage{adjustbox}
\pgfplotsset{compat=1.18}
\usetikzlibrary{pgfplots.colormaps}
\usetikzlibrary{matrix}
\usepackage{multirow} % For multirows
\usepackage{graphicx} % For the rotati
\usepackage{tabularx,booktabs}
\usepackage{diagbox}
\usepackage{pdflscape}
\usepackage{longtable}
\usepackage{array}
\usepackage[table]{xcolor}
\usepackage{tikz}
\usepackage{booktabs}
\usepackage{listings}
\usepackage{ifthen}
\renewcommand{\arraystretch}{1.2}
\usepackage{xcolor}
\usepackage[table]{xcolor}
\usepackage{longtable}
\usepackage{amssymb}
\usepackage{array}
\usepackage{booktabs}
\usepackage{newtxtext,newtxmath}   % Times-like text + math

\usetikzlibrary{positioning, shapes.geometric, arrows.meta}
\usepackage[margin=1in]{geometry}
\newcolumntype{C}{>{\centering\arraybackslash}X}
\theoremstyle{plain}

\theoremstyle{definition}

\theoremstyle{remark}

\usepackage{verbatim}
\usepackage{tablefootnote}
\begin{document}

\title{QuiLL: An LLM-Based Vulnerability Assessment Framework for the Wild}

% \author{Rijha~Safdar, Danyail~Mateen, Syed~Taha~Ali and Wajahat~Hussain  % <-
% \thanks{R. Safdar, S.T. Ali and W. Hussain are with School of Electrical Engineering and Computer Science, National University of Sciences and Technology, Islamabad, Pakistan, 44000. e-mail: rsafdar.dphd19seecs@seecs.edu.pk ,e-mail: taha.ali@seecs.edu.pk, e-mail:wajahat.hussain@seecs.edu.pk}
\author{Rijha~Safdar, Danyail~Mateen, Syed~Taha~Ali and Wajahat~Hussain% <-
\thanks{R. Safdar, S.T. Ali and W. Hussain are with School of Electrical Engineering and Computer Science, National University of Sciences and Technology, Islamabad, Pakistan, 44000. e-mail: rsafdar.dphd19seecs@seecs.edu.pk ,e-mail: taha.ali@seecs.edu.pk}
\thanks{D. Mateen is with the Department
Computer Science, Fast University, Islamabad,
Pakistan, 44000}
}
\maketitle
\begin{abstract}

Large Language Models (LLMs) have demonstrated exceptional progress in multiple domains of software engineering including software vulnerability detection. Using LLMs to automate vulnerability detection in the wild is an important and relatively under-explored problem. In this paper we propose QuiLL, the first comprehensive evaluation framework for real-world vulnerability detection. Our solution consists of an end-to-end pipeline that draws together cutting-edge LLM optimization techniques and strategies specifically catering to the complexities of real-world vulnerability detection. Our specific contributions include (i) diverse prompt designs for vulnerability detection and reasoning (ii) a real-world vector data store constructed from the National Vulnerability Database to provide dynamic in-context learning, and (iii) a novel scoring metric which quantifies accuracy and reasoning quality of model predictions. QuiLL enables researchers to easily and systematically benchmark and compare the vulnerability detection capabilities of various LLMs and assess their readiness for deployment in actual code production pipelines.
\end{abstract}


\IEEEpeerreviewmaketitle

% ABCDEF
%%%%%%%%%%%%
\newcommand{\NLP}{Natural Language Processing }
\newcommand{\DL}{Deep Learning }
\newcommand{\AVD}{Automated Vulnerability Detection}
\newcommand{\NL}{Natural Language}
\newcommand{\PL}{Programming Language}

\newcommand{\bA}{\bar{A}}
\newcommand{\mc}{\mathcal{C}}
\newcommand{\cS}{\mathcal{S}}
\newcommand{\Del}{\Delta}
\newcommand{\del}{\delta}

\newcommand{\E}{\mathbf{E}}
\newcommand{\eps}{\epsilon}
\newcommand{\Gam}{\Gamma}
\newcommand{\gam}{\gamma}
\newcommand{\half}{\frac{1}{2}}

\newcommand{\Sg}{\Sigma}
\newcommand{\sg}{\sigma}

\newcommand{\tA}{\tilde{A}}
\newcommand{\tB}{\tilde{B}}
\newcommand{\tC}{\tilde{C}}

\newcommand{\bx}{\bar{x}}
\newcommand{\hx}{\hat{x}}
\newcommand{\tx}{\tilde{x}}

\newcommand{\by}{\bar{y}}
\newcommand{\ty}{\tilde{y}}
\section{Introduction}
% 1st Paragraph: Software vulnerabilities are out of control - high level numbers

With the rise of digitization and growing deployment of software systems in vital sectors like healthcare, finance, and national security, the issue of software vulnerabilities has assumed critical importance. Software vulnerabilities are growing at an unprecedented rate: in 2024, the National Vulnerability Database (NVD) recorded nearly 40,000 new vulnerabilities, marking the highest annual total to date \cite{nist_nvd_statistics}. Moreover, IBM reported a 10\% increase in the global average cost of a security breach (USD 4.88 million, up from USD 4.45 million in 2023) \cite{ibm_2024}.

% software vulnerabilities - some examples
Exploitable vulnerabilities not only cause significant losses in damages, but can also affect millions of users, and inflict considerable reputational damage to organizations. For instance, the recent Log4Shell vulnerability in Apache Log4j which exposed billions of devices to remote code execution attacks, affected major companies including Apple, Amazon, and Microsoft with estimated average per-incident response cost in excess of USD 90,000 \cite{63}. In another prominent example in 2023, hackers exploited vulnerabilities in MOVEit Transfer, a popular file transfer tool, impacting corporate and government systems, breaching personal data of 77 million people across 2,600 organizations, and causing over USD 9.92 billion in losses \cite{2023_moveit}.

% automating detection - basics + problemslipp2022empirical
These examples make clear the pressing need for effective and scalable automated vulnerability detection solutions. Widely used traditional solutions, which include static analysis, dynamic analysis and hybrid approaches, often miss complex and subtle vulnerabilities. For instance, one empirical study found that state-of-the-art static analyzers missed 47–80\% of vulnerabilities in a benchmark dataset of real-world programs \cite{lipp2022empirical}.

% introducing LLMs
Recently, deep learning solutions (DL), particularly large language models (LLMs), have shown potential to identify vulnerabilities at scale \cite{19},\cite{8}. Tools such as GPT-4, DeepSeekR1, Claude, and Qwen2.5-Coder are transformer-based architectures, pre-trained on large and diverse datasets, and demonstrate exceptional performance on various natural language processing (NLP) tasks. Vulnerability detection may be framed as an NLP task, given the structural similarities to coding, thereby enabling LLMs to detect vulnerabilities, reason about their cause, and create patches \cite{8,34,7,nong2025appatch}.

However, there are significant challenges to be overcome before LLMs can be used for real-world security analysis.

First, \textbf{existing testing methodologies for LLM-based vulnerability detection are largely ad-hoc and dataset-driven}, resulting in evaluations that do not reflect the complexity of real-world software. Various studies find that LLMs demonstrate remarkable results on curated evaluation datasets, but their performance drops significantly on unseen codebases \cite{ding2024vulnerability} \cite{safdar2025data}. For instance, researchers found that state-of-the-art solutions VulDeePecker and SySeVR encounter an average performance drop from $\sim85\%$ to 12.18\% and 10.37\% respectively, when tested on real-world dataset \textsc{Re}\textlarger{V}\textsc{eal} \cite{chakraborty2021deep}. Another study on leading solutions, including LineVul, noted severe performance degradation on new datasets, 
with precision dropping up to 95 percentage points and F1‑scores up to 91 points \cite{8}.

%Chakraborty et al. \cite{chakraborty2021deep} reported that state-of-the-art DL solutions such as VulDeePecker and SySeVR encounter average performance drop from $\sim85\%$ to 12.18\% and 10.37\% respectively, when tested on realistic data \textsc{Re}\textlarger{V}\textsc{eal}. Even after re-training these models on \textsc{Re}\textlarger{V}\textsc{eal} data, their performance drop by from  $\sim85\%$ to 15.7\% and 30.25\% respectively. In another paper of Chakraborty et al. \cite{chakraborty2024revisiting}, they reported that DL vulnerability detectors models including LLMs such as LineVul \cite{8} suffered severe performance degradation, the precision dropped by up to 95 percentage points and F1‑score by up to 91 points, compared to their originally reported results.

% Deep learning - vuldeepecker
% LLM numbers - LineVul
% Third example
% Existing research focuses on either fine-tuning models \cite{7,8} or prompting LLMs\cite{59} for vulnerability detection.

Second, \textbf{the vulnerability detection reasoning workflow suffers from a fundamental flaw in that vulnerabilities are revealed beforehand}. Recent benchmark studies (e.g. \cite{1,34,35,37,38}) evaluate LLMs on predefined, known vulnerabilities, which prevents realistic detection of threats in the wild. In real-world security analysis, neither the presence nor the nature of vulnerabilities is known a priori, a critical constraint that existing benchmarks fail to capture.
  
Likewise, strategies for knowledge augmentation largely rely on synthetic datasets or hand-picked examples, which do not reflect the conditions faced by security analysts in practice. A recent example is a study which uses hand-picked examples (e.g., CWE-190) for few-shot (FS) learning evaluation, querying code instances for that specific security vulnerability \cite{1}. This strategy lacks real-world context and the setting does not depict a real-world scenario.

% \textbf{We need 3-4 lines to show other frameworks are rare/or not applicable.}
% In this paper, we propose \textit{QuiLL}, a standardized testing framework, QuiLL, for LLMs for vulnerability detection in the wild. 

This is also a key limitation of evaluation frameworks proposed in the literature to benchmark the performance of LLMs: these methodologies (e.g. \cite{1},\cite{34},\cite{35}, \cite{9}, \cite{37}, and \cite{38}) address different dimensions of vulnerability detection, but do not provide a comprehensive, real-world evaluation pipeline, which includes real code snippets, real-world context, diverse prompting strategies, automated evaluation, and reasoning-based scoring.

% only align partially with typically address one or two dimension and therefore do not provide a comprehensive, real-world evaluation pipeline
%: most frameworks (e.g. \cite{1},\cite{34},\cite{35}, \cite{36}, \cite{37} and \cite{38}) are limited in scope and only partially align with our objectives, which include real-world code snippets, real-world context, diverse prompting strategies, automated evaluation, and reasoning-based scoring.

%Prior studies typically address one or two dimension and therefore do not provide a comprehensive, real-world evaluation pipeline. Hence, such frameworks are not directly applicable on real-world scenarios. Our benchmark fills this gap by integrating realistic contexts with reasoning-aware metrics and a unified, end-to-end evaluation framework.

% 1-2 lines to justify - clear testing methodology, incorporates newer techniques, reasoning-based scoring, widely applicable

To address this critical gap, we propose a new standardized evaluation framework for LLMs to evaluate their performance in the real world. To promote reproducibility and support future research, we publicly release QuiLL at: \href{https://github.com/Rijha/QuiLL}{QuiLL}. Our precise contributions are as follows:

\begin{enumerate}
\item \textbf{We propose QuiLL, a comprehensive and robust testing methodology and evaluation framework for vulnerability detection, enabling users to compare the performance of LLMs under realistic, ``in the wild'' conditions.} Unlike existing benchmarks in the literature that rely on synthetic datasets with known vulnerabilities, QuiLL operates under a \textit{fair evaluation paradigm} where models must detect and reason about software vulnerabilities in realistic settings without prior knowledge of their existence or type.

%QuiLL specifically caters to vulnerability detection in the wild, providing users a clear and comprehensive testing methodology to compare the performance of different LLMs.

% QuiLL integrates realistic contexts with reasoning-aware metrics, and incorporates cutting-edge innovations, including newer prompting strategies, reasoning-based scoring, and retrieval-augmented generation, to ensure wide applicability in real-world settings.

\item  \textbf{We propose a novel evaluation scoring metric that combines both the accuracy of the model's prediction and the quality of it's reasoning}. This metric addresses a critical gap in existing evaluation methodologies, which have thus far considered prediction accuracy and reasoning independently of each other. However, in security-critical contexts, correct predictions with flawed reasoning have little actionable value and may even prove dangerously misleading, providing false confidence with serious ramifications. Our metric combines binary prediction accuracy with cosine similarity-based semantic reasoning, thereby offering a holistic, automated, and scalable measure of the model's performance. Through systematic analysis, we also identify optimal prompting strategies for practical vulnerability detection and reasoning which significantly enhance model performance.

\item \textbf{To enable real-time retrieval-augmented generation (RAG), we build a comprehensive, real-world vectorized knowledge base data store from the NVD}, transforming its CVE entries, patch codes, and commit metadata into high-dimensional embeddings. This vectorized store enables semantic similarity search and real-time RAG. Unlike static, hand-picked examples used in prior work \cite{1}, our work enables dynamic in-context learning with high-impact real-world vulnerabilities, thereby providing deeper insights about LLM performance under realistic, security-critical conditions. This is a rich dataset, approximately 50 GB in size, and comprising 12,107 vulnerability-fixing commits, covering 11,873 unique CVEs across 272 distinct CWE categories, extracted from 4,249 open-source projects.

\end{enumerate}

To the best of our knowledge, QuiLL represents the first evaluation framework specifically catering to the complexities of real-world vulnerability detection. We validate our evaluation framework in experiments using five diverse state-of-the-art models: GPT-4, Gemini-1.5-Flash-002, Qwen2.5-Coder-14B, LLaMA-3-8B and Phi-4. We employ four structured prompting paradigms: standard, chain-of-thought, decomposition, and plan-and-solve. We test for 15 real-world critical vulnerabilities from MITRE's Top 25 CWEs and their corresponding patches, totaling 1,200 evaluation instances. Results indicate that QuiLL provides a rigorous testbed that reflects the nuance and diversity of real-world security analysis.

% snapshot of results
Our results also confirm and expand upon several insights from the research literature. Detection is poor in zero-shot settings and incorrect reasoning is a critical bottleneck. CVE-based retrieval leads to significant improvements across all models (4-22\%), particularly for closed-source models (GPT-4 and Gemini). We also observe a significant overlap in vulnerabilities detected by different models, which suggest a pronounced lack of diversity in training data and the tendency of models to generalize on familiar patterns.

% snapshot II
Structured prompts (such as decomposition and plan-and-solve) also significantly enhance detection performance by 6-10\% on average and substantially more for larger models, and also lead to more coherent and accurate reasoning. Bigger is also clearly better: models with a larger parameter count show markedly better performance. Moreover, our novel scoring metric effectively differentiates between correct predictions and reasoning accuracy. This feature has significant utility in follow-on applications such as vulnerability localization and patching.

% snapshot III and our contibution
Most important, however we confirm that despite their promise, advanced LLMs still struggle under real-world conditions and are not yet ready for large-scale deployment. These findings need not be interpreted negatively but rather indicate immense opportunity for future research. We hope the QuiLL evaluation framework provides researchers with an intuitive and structured method innovate and validate their findings.

The rest of this paper is organized as follows: Section II provides essential background on automated vulnerability detection and prior work. Section III describes our real-world evaluation framework QuiLL. Section IV presents the results of our experiments with leading LLMs followed by discussion of key insights in Section V. We conclude in Section VI with directions for future research.

%  HARMONIZE: In vulnerability detection, an accurate prediction without correct reasoning can be misleading. Existing metrics evaluation is based on either prediction or reasoning independently, overlooking a combined scoring metrics. A model that correctly identifies code as vulnerable but attributes the flaw to an incorrect root cause provides little actionable value to security practitioners. This motivated us to design a hybrid metric that jointly evaluates the correctness of vulnerability detection and the quality of reasoning, offering a more realistic and comprehensive assessment.

%Background
\section{Background and Prior Work}

%summary of this section
Here we summarize the evolution of software vulnerability detection followed by a comprehensive review of evaluation frameworks in the literature.

%The detection of software vulnerabilities has evolved from traditional methods to AI-driven approaches. This section provides a comprehensive review of vulnerability detection solutions and frameworks.

\subsection{Automated Vulnerability Detection}

%Traditional approaches
Traditional vulnerability detection techniques comprise static, dynamic, and hybrid analysis.

\textbf{Static analysis tools} analyze raw source code for errors. These tools are widely used in industry but suffer from critical limitations including the inability to interpret complex control and data flows, high false positive rates, and poor performance \cite{42,43,44}. One study indicates that state-of-the-art static analysis tools miss 47–80\% of known real-world vulnerabilities \cite{lipp2022empirical}. In contrast, \textbf{dynamic analysis tools} capture security issues during runtime \cite{45}. However, this process is computationally expensive, time-consuming, and usually requires sandboxed environments and security experts to operate \cite{baldoni2018survey}, \cite{kirda2006behavior}.

\textbf{Hybrid analysis} combines static and dynamic approaches, resulting in improved performance \cite{46}. Shields et al. demonstrate that hybrid analysis increases accuracy by 17\% with upto 25\% faster detection \cite{shields2023hybrid}. However, this approach also demands extensive resources, manual effort, and domain expertise.

%DL based approaches
Researchers are investigating machine learning solutions to overcome the limitations of traditional approaches. This includes \textbf{deep learning techniques}, which incorporate control and data flow-based representations, constructed by tools like Joern using Code Property Graphs \cite{31}. These approaches demonstrate effective results: for example, VulDeePecker achieved F1-scores exceeding 90\% on C/C++ vulnerabilities \cite{19}.

%LLM based approaches
The rise of \textbf{large language models} promises to fundamentally change the vulnerability detection landscape. For instance, LineVul, a transformer-based model using pre-trained language representations, achieves F1-scores up to 0.91 on real-world C/C++ datasets \cite{7}. Nong et al. propose APPATCH, which uses adaptive prompting to patch real-world applications, resulting in up to 60\% correct patches for unseen vulnerabilities \cite{nong2025appatch}. 

%ML and DL challenges
However, machine learning approaches face considerable challenges: VulDeePecker and SySeVR achieve high accuracy on curated datasets but exhibit major generalization failures in real-world settings, with accuracy plummeting on unseen projects ($\approx10-12\%$) \cite{chakraborty2021deep}. Ullah et al. report that LLMs like GPT-4 and PaLM-2 misclassify vulnerabilities in 17–26\% of cases and frequently fail to distinguish between vulnerable and patched code. LLMs also fail to assess semantic equivalence: mere variable renaming can trigger wrong assessments \cite{1}. Khare et al. demonstrate that LLMs struggle with context-dependent, multi-function vulnerabilities, achieving only 40–60\% F1-scores and major performance degradation on unseen vulnerability types \cite{34}. For details of these challenges, we refer the reader to a comprehensive survey \cite{sheng2025llms}.

These findings highlight three key points: first, that current machine learning solutions are dataset dependent which limits their use in industry. Second, there is a distinction between an LLM's ability to detect vulnerabilities and to reason about them correctly. And finally, these findings emphasize the need for a sophisticated and robust evaluation methodology to assess vulnerability detection for real-world applications.

\subsection{Model Adaptation Techniques}

%fine-tuning and prompting
Important strategies have emerged to adapt LLMs for vulnerability detection, including fine-tuning, prompting and retrieval-augmented generation. In the first approach, \textbf{fine-tuning}, a pre-trained model is fine-tuned on a vulnerability-specific dataset to further improve its performance. Safdar et al. demonstrated that fine-tuning models using high quality and diverse real-world datasets significantly improves their ability to generalize \cite{safdar2025data}. For the BigVul benchmark dataset, they demonstrated 6.8\% improvement in recall.

The second approach, \textbf{prompt-based inference}, leverages zero-shot or few-shot learning through carefully crafted prompts \cite{1,9}. Zheng et al. demonstrated that structured prompts yield substantial improvements over simple prompting by enforcing intermediate reasoning steps \cite{25}. Likewise, Sun et al. reported that step-by step prompting improves GPT-4’s accuracy 30-47\% at identifying vulnerabilities \cite{9}.

Another promising optimization, \textbf{retrieval-augmented generation (RAG)} provides LLMs contextual domain-specific information to mitigate knowledge limitations and hallucination tendencies. RAG can significantly improve vulnerability detection: Du et al. proposed Vul-RAG, which achieves 16–24\% accuracy improvements over zero-shot approaches \cite{64}. Liu et~al. evaluated a range of LLMs using knowledge augmentation and identified 14 zero-day vulnerabilities in bug bounty programs \cite{37}.

These optimizations yield significant performance returns and it is essential to incorporate them in testing methodologies and evaluation frameworks for improved vulnerability detection.

\subsection{Prior Work: Evaluation Frameworks}

Here we describe key evaluation studies which benchmark the performance of LLMs for vulnerability detection.

Khare et al. undertake \textbf{a systematic evaluation of 16 pre-trained LLMs} including GPT-4 and CodeLLaMA variants on synthetic and real-world C/C++ and Java code \cite{34}. They report an average accuracy of $\approx62.8\%$, F1 score of $\approx0.71$, and markedly better performance detecting simple bugs such as OS command injection and null pointer dereference vulnerabilities. Moreover, they demonstrate that advanced prompting strategies like step-by-step analysis can significantly improve F1 scores by up to $\approx 0.18$. This finding motivates us to adopt step-by-step reasoning prompts in our own evaluation framework.

%Gao et al. introduce \textbf{\emph{VulBench} which combines high-quality data from Capture-the-Flag (CTF) challenges and real-world applications} to evaluate multiple LLMs \cite{35}. Their results again highlight the effectiveness of LLMs at detecting simple vulnerabilities, notable performance variation across datasets, and significant performance degradation when addressing more complex and real-world vulnerability patterns.

Gao et al. introduce \textbf{\emph{VulBench}}, which  evaluates multiple LLMs across multiple datasets, including Juliet C/C++, OpenSSL, Big-Vul, and Devign \cite{35}. The results show that exisiting models achive notable performance i.e., 80\% on vulnerability identification and classification tasks, but performance degrades to less than 30\% accuracy on more detailed vulnerability analysis tasks.

Sun et al. propose \textbf{\textit{LLM4Vuln}, a unified reasoning-based evaluation framework to understand and identify vulnerabilities in Solidity-based smart contracts} \cite{9}. This investigation attempts to decompose the vulnerability detection task into multiple sub-tasks (e.g. vulnerabiilty reasoning, additional information seeking, context supplementation, and knowledge enhancement). Significant variation in results is observed due to these steps, particularly with GPT-4, with detection accuracy improving from approximately 30\% to 47\% for certain vulnerabilities, as well as detection of multiple zero-day vulnerabilities in bug bounty programs. 

Zibaeirad et al.~\cite{38} introduce \textbf{\emph{VulnLLMEval}}, a framework designed to evaluate LLMs for vulnerability detection and patching. They experimented with 307 vulnerabilities extracted from the Linux kernel, covering \textbf{285 CVEs across 30 CWE categories} and providing paired vulnerable and patched code samples. Their results emphasized key limitation of LLMs that it often struggle to differentiate between vulnerable and patched code. 

%Zibaeirad et al. \cite{38} and Ullah et al. \cite{1} both evaluated LLMs and emphasized key limitation of LLMs that it often struggle to differentiate between vulnerable and patched code.

Ullah et al. \cite{1} present \textbf{\emph{SecLLMHolmes}}, a detailed evaluation framework that compares eight state-of-the-art LLMs across multiple investigative dimensions on 228 deliberately crafted code scenarios \cite{1}. They demonstrate that LLMs often provide non-deterministic results, exhibit incorrect reasoning, and perform poorly on real-world detection tasks. For example, even advanced models like PaLM 2 and GPT-4 make incorrect assessments in 26\% and 17\% of cases respectively by merely adding library function in the source code or by changing function/variable names.These findings highlight that current LLMs are not ready for automated vulnerability identification and reasoning in realistic settings.

These evaluation frameworks include rigorous methodologies, automated pipelines and diverse metrics, but critical gaps exist, which we seek to address in our work. First and foremost, these assessments primarily rely on known vulnerabilities, synthetic data, and hand-crafted context, which fails to reflect real-world application scenarios, where the presence or the nature of vulnerabilities is not explicitly known beforehand.

Second, these frameworks do not include dynamic real-world context retrieval, which have been demonstrated to significantly improve detection accuracy. Moreover, we find that most of these solutions overlook or do not adequately differentiate between the detection and reasoning capabilities of LLMs, which can have significant implications on the utility or actionable value of the model predictions. %SecLLMHolmes addresses this concern using majority voting mechanism based on ROUGE scores 

%Also, their reliance on majority voting using ROUGE scores evaluations raises concern. ROUGE (overlapping words, not correct reasoning or semantic understanding) based n-gram matching fails to evaluates semantic reasoning and hence it is not a reliable measure for strong evaluation. 

%Despite recent progress, vulnerability detection remains unreliable in real-world settings. LLM-based vulnerability detection struggle to distinguish vulnerable from patched code \cite{1}, \cite{38}, and generalize poorly to unseen vulnerabilities, with performance typically limited to 40–60\% F1 even under RAG augmentation \cite{34,37}. 

%\begin{figure}[t!]
%    \centering
%    \includegraphics[width=0.49\textwidth, height=7cm]{RW_new.png} 
%    \captionsetup{font=small}
%    \caption{Related work in Software Vulnerability Detection}
%    \label{fig:RW}
%\end{figure}

%Some work \cite{9,37,38} utilizes real-world data but others \cite{1,34,35} only partially utilizes such data, often relying on synthetic datasets that do not reflect the actual complexity found in the wild. 

To address these shortcomings we next introduce QuiLL, our automated, real-world, in-context evaluation framework. QuiLL uses diverse prompting strategies and a rich array of metrics. Our framework systematically assesses LLM-based vulnerability detection on previously unseen code augmented with real-world context, enabling a more faithful and rigorous evaluation of model reliability in complex real-world conditions.

%Motivation and Scope
%With the rapid progress of LLM capabilities, hundreds of models released over the past four years \cite{60} signifying the urgent need for standardized, real-world benchmarking frameworks to assess their effectiveness in security-critical tasks. This motivates us to design a real-world in-context based evaluation framework QuiLL,to assess capability of LLMs for vulnerability detection and reason of vulnerability. Also, neither the presence nor the nature of a vulnerability is known a priori unlike prior work\cite{1}. This closely reflects practical scenario faced by a security analysts. We focus on C/C++ vulnerabilities—languages that dominate critical infrastructure software and leverage comprehensive, authentic vulnerability data from the National Vulnerability Database (NVD) \cite{15} to enable dynamic, retrieval-augmented in-context learning.

%While our current evaluation focuses on C/C++ and belongs to  MITRE's Top 25 Most Dangerous Software Weaknesses, the QuiLL framework can be extended to other programming languages, making it broadly applicable across diverse applications. The advancement in software vulnerability detection is summarized in Fig. \ref{fig:RW}.

%Our approach addressesthe prior limitation by incorporating real-world in-context based evaluation on real-world data for vulnerability detection, supporting diverse prompting strategies, and deploying automated, multidimensional evaluation (zero-shot and in-context). This framework also incorporates a metrics integration to evaluate both correctness. 

\section{The QuiLL Framework}

%We now describe QuiLL, our evaluation framework to assess LLMs vulnerability detection and reasoning capability in ``in-the-wild'' (real-world) conditions. 
% Unlike prior work that primarily relies on binary classification accuracy or handcrafted examples, QuiLL integrates structured prompting strategies, retrieval-augmented contextual knowledge, and automated evaluation to provide a holistic assessment of LLM performance on real-world vulnerability detection tasks.

\begin{figure}[t!]
    \centering
    \includegraphics[width=\columnwidth]{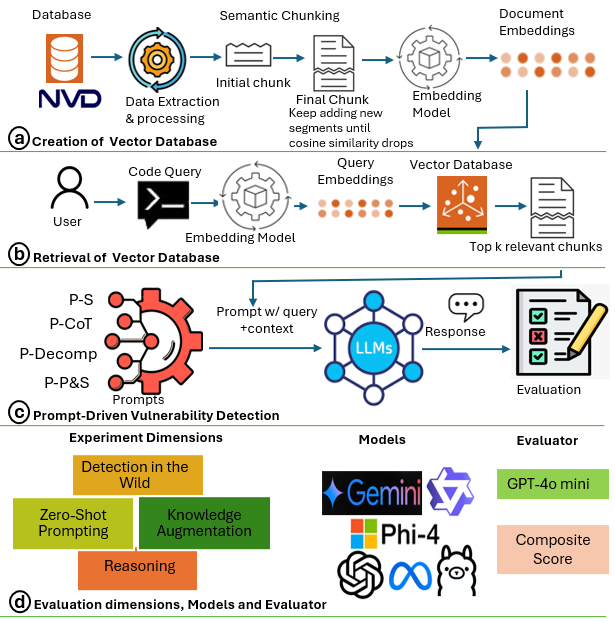}
    \captionsetup{font=small}
    \caption{
Overview of the QuiLL framework.} 
% \protect\textcircled{a} Creation of a vectorized vulnerability knowledge base from NVD data. 
% \protect\textcircled{b} Retrieval of semantically similar vulnerability examples using embedding search. 
% \protect\textcircled{c} Prompt-driven vulnerability detection using LLMs with multiple prompting strategies. 
% \protect\textcircled{d} Automated evaluation across experimental dimensions including prediction accuracy and reasoning quality.}
    \label{fig:Hawkllm}
\end{figure}

QuiLL's workflow consists of four distinct stages as depicted in Fig.~\ref{fig:Hawkllm}: 
\textcircled{a} construction of a vectorized vulnerability knowledge base, 
\textcircled{b} retrieval of relevant contextual examples, 
\textcircled{c} prompt-driven vulnerability detection, and 
\textcircled{d} automated evaluation across multiple experimental dimensions. 

We next describe these stages in detail.
\definecolor{bgcolor}{RGB}{255,255,255}        % White background
\definecolor{codegray}{gray}{0.1}              % Dark text
\definecolor{keywordblue}{rgb}{0,0,0.8}        % Blue keywords
\definecolor{commentgreen}{rgb}{0,0.5,0}       % Green comments
\definecolor{stringpurple}{rgb}{0.6,0,0.6}     % Purple strings

\lstdefinestyle{custompython}{
    backgroundcolor=\color{bgcolor},
    basicstyle=\ttfamily\footnotesize\color{codegray},      % Monospaced, not italic
    keywordstyle=\color{keywordblue}\bfseries,              % Bold blue keywords
    commentstyle=\color{commentgreen},                      % Green comments (no \itshape)
    stringstyle=\color{stringpurple},                       % Purple strings
    breaklines=true,
    frame=single,
    showstringspaces=false,
    captionpos=b,
    language=Python,
    aboveskip=0pt,
    belowskip=0pt,
    xleftmargin=0pt,
    xrightmargin=0pt,
    lineskip=-1pt
}

\begin{figure}[!t]
\centering
\begin{lstlisting}[style=custompython]
query = f"""
--DESCRIPTION--
         This code is part of{"project name"} project, written in{"language"} and has{"CWE"}, which means{"concise description"}.In the following code{"description"}
--/DESCRIPTION--
Vulnerable Code:
                    {CODE}
Fixed Code:
                    {PATCH}
--COMMITMSG--
                    {commit msg}
--/COMMITMSG--"""
\end{lstlisting}
\captionsetup{font=small}
\caption{Structured prompt template used for querying LLMs}
\label{fig:query}
\end{figure}

\definecolor{lightblue}{RGB}{232,240,254}    % ZS
\definecolor{lightcyan}{RGB}{220,255,255}    % ZS-C
\definecolor{lightyellow}{RGB}{255,255,220}  % Semi Wild-DB
\definecolor{lightorange}{RGB}{255,235,200}  % Wild-DS
\definecolor{lightgreen}{RGB}{230,255,230}   % Role-KA
\definecolor{lightpink}{RGB}{255,230,240}    % Role-KA-C

\lstdefinestyle{custompython}{
    backgroundcolor=\color{bgcolor},
    basicstyle=\ttfamily\footnotesize\color{codegray},      % Monospaced, not italic
    keywordstyle=\color{keywordblue}\bfseries,              % Bold blue keywords
    commentstyle=\color{commentgreen},                      % Green comments (no \itshape)
    stringstyle=\color{stringpurple},                       % Purple strings
    breaklines=true,
    frame=single,
    showstringspaces=false,
    captionpos=b,
    language=Python,
    aboveskip=0pt,
    belowskip=0pt,
    xleftmargin=0pt,
    xrightmargin=0pt,
    lineskip=-1pt
}
\begin{figure}[!t]
\centering
\begin{lstlisting}[style=custompython]
sys = """You are a security expert. Answer the following question clearly and concisely. For your reference we have provided a similar example. The following example is structured as Description, Vulnerable Code, Fixed Code and finally Code to be evaluated. You just need to provide a concise answer containing prediction and reasoning of "Code to be evaluated".
Context:\n{context}. """
\end{lstlisting}

\captionsetup{font=small}
\caption{System prompt incorporating retrieved examples for LLM guidance.}
\label{fig:sys_prompt}
\end{figure}
%dataset
\subsection{Creation of Vector Database}
In the first stage \textcircled{a} we construct a vectorized knowledge base using the publicly available \texttt{CVEfixes} v1.0.8 dataset \cite{14}. CVEfixes mines public Common Vulnerabilities and Exposures (CVE) records from the public US National Vulnerability Database (NVD) to create a linked dataset at the commit, file, and method levels.

%\footnote{For knowledge augmentation, we leveraged the real-world publicly available \texttt{CVEfixes} dataset\cite{14}, version 1.0.8, which includes all published CVEs up to July 2024, comprising SQL dump of approximately 50 GB in size. It includes rich metadata such as vulnerable and patch codes and their descriptions which can also be used for training vulnerability patching, vulnerable line and cwe classification. In total, the dataset comprises 12,107 vulnerability-fixing commits, covering 11,873 unique CVEs across 272 distinct CWE categories, extracted from 4,249 open-source projects.}

This dataset is an SQL dump comprising all published CVEs up to July, 2024. It includes rich metadata, including vulnerable and patch codes and their descriptions. In total, the dataset, amounting to approximately 50 GB in size, comprises 12,107 vulnerability-fixing commits, covering 11,873 unique CVEs across 272 distinct CWE categories, extracted from 4,249 open-source projects.

The structured format of this dataset poses a challenge in that it is designed primarily for human browsing and keyword-based search and cannot be directly leveraged by LLMs for in-context reasoning. We resolve this issue by transforming CVE entries, patch codes, and commit metadata into high-dimensional vector embeddings, thereby enabling semantic similarity search and retrieval-augmented generation. 

To build the vector data store, we use Qdrant~\cite{qdrant2026}, an open-source free vector database. Qdrant is optimized for handling high-dimensional embeddings, efficient querying, indexing, and context retrieval. We generate the embeddings using OpenAI’s \texttt{text-embedding-3-small} model, released in January 2024~\cite{16}. This model is widely adopted in large-scale RAG pipelines and frameworks, including LangChain~\cite{66} and LLaMAIndex~\cite{67}. Featuring performance comparable to large models~\cite{65}, \texttt{text-embedding-3-small} generates 1,536 dimensional embeddings, supports a maximum context length of 8,191 tokens, and is cost-efficient at \$0.02 per one million tokens. 

To ensure semantically coherent code segments and descriptions, we use LangChain's Semantic Chunker, backed by the same embedding model. These embeddings are stored in a Qdrant database along with structured payloads, enabling retrieval of fine-grained semantically similar examples at run-time. Fig.~\ref{fig:query} depicts the template we use for in-context examples. These samples are then appended to structured prompts, as highlighted in Fig.~\ref{fig:sys_prompt}, to aid LLMs in vulnerability detection and reasoning.

%We utilize this vector database and associated metadata to construct meaningful in-context examples as per the template depicted in Fig.~\ref{fig:query}

%A structured prompt as shown in Fig.\ref{fig:sys_prompt} is used to aid in vulnerability detection and reasoning.
% Unlike prior work that relies on binary accuracy only or hand-crafted examples, QuiLL integrates three features: (i) simple to structured prompting strategies (standard, chain-of-thought, decomposition, and plan-and-solve) to guide step-wise reasoning, (ii) an automated evaluator that jointly measures prediction accuracy and semantic reasoning quality, and (iii) a vectorized knowledge base enabling dynamic, retrieval-augmented context provision based on real-world vulnerability data. 

\subsection{Retrieval-Augmented Generation}
In the second stage \textcircled{b} relevant contextual examples are retrived from the vector database to supplement and guide the LLM model's reasoning capacity. During inference, the input code sample is first converted into an embedding and a similarity search is undertaken in the vector database to retrieve the top-\textit{k} semantically similar examples, including their descriptions, vulnerable code, and corresponding patches. These retrieved examples are incorporated into the LLM prompt as contextual references. We set \textit{k}=5, following \cite{hasan2025llm}, where k = 5 was chosen as it provides optimal accuracy with minimal noise, since larger k values show diminishing returns and reduced efficiency.

\subsection{Prompt-Driven Vulnerability Detection }
In the third stage \textcircled{c} the retrieved contextual examples are combined with structured prompts to evaluate the detection and reasoning capabilities of LLMs.

Prompting strategies include Standard prompts (P-S), Chain-of-Thought (P-CoT), Decomposition (P-Decomp), and Plan-and-Solve (P-P\&S). Each strategy incrementally guides the reasoning process of LLMs with structured, multi-step reasoning, in an attempt to approximate the behavior of security experts working in real-world settings.

While P-S, P-CoT,P-Decomp and P-P\&S prompting strategies have been widely explored in prior literature \cite{kojima2022large},\cite{1}, \cite{khot2022decomposed}, \cite{vatsal2024survey}, they have not been thoroughly evaluated for zero-shot and in-context learning in real-world vulnerability detection settings. 

The details of each strategy are as follows:
\begin{itemize}

\item \textbf{Standard Prompt (P-S)}: this is a baseline prompt that asks the model to classify code as vulnerable or secure. This minimal prompt measures the model’s inherent reasoning ability without additional guidance. It helps establish a reference point for evaluating the benefits of more structured prompting.

\item \textbf{Chain-of-Thought Prompt (P-CoT)}: this strategy extends the standard prompt by encouraging the model to reason step by step using phrases such as ``Let's think step by step''. Kojima et al. demonstrated that P-CoT can dramatically improve zero-shot reasoning without additional training data, for instance, increasing accuracy on the MultiArith benchmark from 17.7\% to 78.7\% \cite{kojima2022large}. In the case of vulnerability detection, P-CoT helps the model explicitly trace program logic, identify unsafe API calls, and explain how insecure data flows may lead to exploitable conditions.

\begin{figure}[t!]
    \centering
     \includegraphics[width=0.8\linewidth]{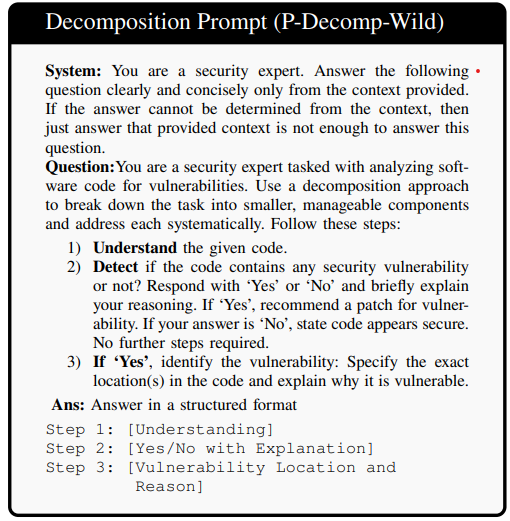}
    \captionsetup{font=small}
    \caption{An example of decomposition prompting where a task into subtasks and structured steps to aid LLM before reaching a conclusion}
    \label{fig:Decomp}
\end{figure}
\item \textbf{Decomposition Prompt (P-Decomp)}: this prompt breaks a task into subtasks and structured steps to aid in understanding, locating, and reasoning about vulnerable and secure code samples before reaching a conclusion. An example of decomposition prompting is provided in Fig.~\ref{fig:Decomp}.

% \begin{tcolorbox}[colback=gray!5!white,colframe=black,title={Decomposition Prompt (P-Decomp-Wild)}]
% \label{lst:p-decomp}
% \scriptsize{
% \textbf{System:} You are a security expert. Answer the following question clearly and concisely only from the context provided. If the answer cannot be determined from the context, then just answer that provided context is not enough to answer this question.

% \textbf{Question:}You are a security expert tasked with analyzing software code for vulnerabilities. Use a decomposition approach to break down the task into smaller, manageable components and address each systematically. Follow these steps:
% \begin{enumerate}
%     \item \textbf{Understand} the given code.
%     \item \textbf{Detect} if the code contains any security vulnerability or not? Respond with `Yes' or `No' and briefly explain your reasoning.
%     If `Yes', recommend a patch for vulnerability.
%     If your answer is `No', state code appears secure. No further steps required.
%     \item \textbf{If `Yes'}, identify the vulnerability: Specify the exact location(s) in the code and explain why it is vulnerable.
% \end{enumerate}

% \textbf{ Ans:} Answer in a structured format
% }
% \begin{verbatim}
% Step 1: [Understanding]
% Step 2: [Yes/No with Explanation]
% Step 3: [Vulnerability Location and
%          Reason]
% \end{verbatim}
% \end{tcolorbox}

This strategy is inspired by multi-step task decomposition, which promotes systematic reasoning and interpretation, thereby mitigating errors due to overly direct or heuristic responses, and more closely approximating the analytical reasoning process of security experts \cite{dua2022successive,press2022measuring}. Khot et al. demonstrate that modular task decomposition leads to a 17-point improvement on MultiArith (78\% to 95\%) and a 14-point improvement on GSM8K (36\% to 50.6\%) compared to standard P-CoT \cite{khot2022decomposed}.

\item \textbf{Plan-and-Solve Prompt (P-P\&S)}: this strategy, proposed by Wang et al., involves two explicit stages \cite{27}: first, in the \textit{planning phase}, the LLM is tasked with creating a high-level plan outlining how it intends to approach the problem. Then, in the \textit{solving phase}, it executes that plan step by step to formulate a conclusion. This approach enhances logical reasoning and helps improve responses. For instance, Vatsal et al. report that P-P\&S improves accuracy over P-CoT by at least 5\% over almost all datasets in the Mathematical Problem Solving task in a zero-shot setting \cite{vatsal2024survey}. 

In the case of vulnerability detection, P-P\&S prompting allows the model to outline a structured approach (e.g., “first check input validation, then memory handling, then privilege escalation risks”) before applying it to the given code sample. 
\end{itemize}
 
We evaluate these prompting strategies for both zero-shot and few-shot learning as detailed in Table \ref{tab:prompt}, enabling us to precisely determine the contribution of structured prompting and context retrieval towards vulnerability detection. The enriched prompts are passed to representative large language models. An illustrative example of this process is shown in Fig.~\ref{fig:llm-prompt-response}.

\begin{table}[t]
\centering
\scriptsize
\setlength{\tabcolsep}{2pt}
\renewcommand{\arraystretch}{1.05}
% \begin{tabular}{p{1.5cm}p{5.7cm}}
\begin{tabular}{>{\centering\arraybackslash}p{1.5cm}
                >{\centering\arraybackslash}p{5.7cm}}
\toprule
\textbf{Prompt ID}   & \textbf{Description} \\
\midrule
\midrule
\multicolumn{2}{l}{\textbf{\textit{Standard Prompts(P-S)}}} \\
\midrule
\textbf{P-S-ZS}   & Security expert role. Predict and reason whether the code instance is \textbf{vulnerable or secure}. if vulnerable inquire about CWE.\\ 

\textbf{P-S-FS}   & Same as P-S-ZS and fetch examples from RAG vector database.\\
\midrule
\multicolumn{2}{l}{\textbf{\textit{Chain-of-Thought Prompts (P-CoT)}}} \\
\midrule
\textbf{P-CoT-ZS} & To instruct LLM to explain reasoning add “Let’s think step by step” is added to P-S-ZS. \\
\textbf{P-CoT-FS}  & Same as P-CoT-ZS and fetch examples from RAG vector database.\\
\midrule
\multicolumn{2}{l}{\textbf{\textit{Decomposition Prompts (P-Decomp)}}} \\
\midrule
\textbf{P\-Decomp-ZS} & Structured steps to understand, detect, locate and reason about \textbf{CWE}.\\
\textbf{P\-Decomp-FS} & Same as P-Decomp-ZS and fetch examples from RAG vector database.\\
\midrule
\multicolumn{2}{l}{\textbf{\textit{Plan-and-Solve Prompts (P-P\&S)}}} \\
\midrule
\textbf{P-P\&S-ZS} & Add plan and "Let's first understand the problem and devise a plan to solve the problem. Then, let's carry out the plan and solve the problem step by step" to P-S-ZS \\
\textbf{P-P\&S-FS} & Same as P-S-ZS and fetch examples from RAG vector database.\\

\bottomrule
\end{tabular}
\caption{Prompt Templates for Different Strategies.}
\label{tab:prompt}
\end{table}

\definecolor{promptblue}{RGB}{180,210,255}
\definecolor{responseyellow}{RGB}{255,250,210}
\definecolor{codebg}{RGB}{240,240,240}
\definecolor{arrowgray}{RGB}{80,80,80}

% Code block style
\lstdefinestyle{mystyle}{
  backgroundcolor=\color{codebg},
  commentstyle=\color{gray},
  keywordstyle=\color{blue},
  numberstyle=\tiny\color{gray},
  stringstyle=\color{red!70!black},
  basicstyle=\ttfamily\footnotesize,
  breaklines=true,
  breakatwhitespace=true,
  tabsize=2,
  showstringspaces=false
}
\lstset{style=mystyle}

\begin{figure}[t!]
\centering
\small
% SYSTEM PROMPT BOX
% \tcbset{colback=promptblue, colframe=blue!80!black, boxrule=0.5pt, arc=3pt, left=5pt, right=5pt, top=4pt, bottom=4pt}
\tcbset{
  colback=promptblue, colframe=blue!80!black,
  boxrule=0.4pt, arc=2pt,
  left=3pt, right=3pt, top=2pt, bottom=2pt,
  fonttitle=\bfseries\footnotesize
}
\begin{tcolorbox}[title=\textbf{System Prompt to LLM}]
\scriptsize
{ You are a security expert. Answer the following question clearly and concisely.  
For your reference we have provided a similar example.
The following example is structured as Description, 
Vulnerable Code, Fixed Code and finally Code to be evaluated.
You just need to provide prediction and reasoning of Code to be evaluated. If the code is vulnerable then identify potential CWE as well.}
\end{tcolorbox}
\vspace{-1ex}
% ARROW
\tikz{\draw[->, thick, color=arrowgray] (0,0) -- (0,-0.4);}

% % QUESTION + CODE BOX
% \tcbset{colback=codebg, colframe=gray!80!black, boxrule=0.5pt, arc=3pt, left=5pt, right=5pt, top=4pt, bottom=4pt}
% QUESTION + CODE BOX
\tcbset{
  colback=codebg, colframe=gray!80!black,
  boxrule=0.4pt, arc=2pt,
  left=3pt, right=3pt, top=2pt, bottom=2pt
}
\begin{tcolorbox}[title=\textbf{User Prompt (Question + Code)}]
\scriptsize{ 
\textbf{Question:}  
Does this code contain instances of security vulnerability or not?
Format your response strictly as:  
\texttt{Prediction: [Yes/No]}  
\texttt{Reason: [Explain in under 50 words]}  
}
\textbf{Code:}
\begin{lstlisting}[language=C]
void copy_input(char *src) {
    char buf[10];
    strcpy(buf, src); }// Potential overflow
\end{lstlisting}
\end{tcolorbox}
\vspace{-1ex}
% ARROW
\tikz{\draw[->, thick, color=arrowgray] (0,0) -- (0,-0.4);}

% RESPONSE BOX
% \tcbset{colback=responseyellow, colframe=orange!70!black, boxrule=0.5pt, arc=3pt, left=5pt, right=5pt, top=4pt, bottom=4pt}

\tcbset{
  colback=responseyellow, colframe=orange!70!black,
  boxrule=0.4pt, arc=2pt,
  left=3pt, right=3pt, top=2pt, bottom=2pt
}
\begin{tcolorbox}[title=\textbf{LLM Response}]
\scriptsize{
\textbf{Prediction:} Yes

\textbf{Reason:} The code uses \texttt{strcpy()} without checking the input length, which can cause a buffer overflow. This is a classic case of CWE-121: Stack-Based Buffer Overflow.
}
\end{tcolorbox}

\caption{Prompt Template: The LLM is provided with a structured instruction, a system prompt assigning role of security expert, a user-defined query, and a real-world code snippet. It outputs prediction and reason. Prediction identifies the vulnerability and reason present the cause/justification.}
\label{fig:llm-prompt-response}
\end{figure}

%To enable few-shot–like behavior, the model dynamically augments its context with real-world vulnerability–patch examples. CVE metadata is obtained from the NVD, while corresponding patch information is collected from linked vendor advisories and public code repositories. These vulnerability–patch pairs are embedded and stored in a vector database. At inference time, semantically similar examples are retrieved based on the input query and incorporated into the model’s prompt, providing task-relevant contextual grounding.

% The context for few-shot is provided using real-world CVEs and their corresponding patched samples. They are collected and transformed into a vectorized knowledge base using embeddings and stored in a vector database. This enables semantic retrieval of vulnerability–patch pairs. To analyze a code sample, the retriever module fetches semantically similar examples from the vector store and it is provided as additional context. In the following subsections, we describe each stage of the pipeline in detail, following the flow shown in Fig.~\ref{fig:Hawkllm}.

%picture of framework

\subsection{Evaluation}

In the final stage \textcircled{d}, our evaluation pipeline assesses the output of the models using binary prediction accuracy (correct/incorrect detection) and reasoning quality. This enables automated, consistent, and scalable evaluation under realistic and practical scenarios.

Our approach draws inspiration from recent work such as MT-Bench \cite{69} and KoalaEval \cite{70} which assess the quality of LLM outputs using semantic comparison based on embeddings. Our approach is in contrast to existing methodologies which largely focus on binary correctness. A model that correctly identifies code as vulnerable and also correctly identifies vulnerability location and CWE type provides significantly more actionable value.
 
%To account for overall model performance, we use accuracy that measures whether the LLM correctly predicts if the code is vulnerable or secure, cosine similarity measures how semantically close the LLMs reasoning explanation is to the ground truth (GT). 

Given an LLM response, we use gpt-4o-mini with a structured prompt to extract: (i) a binary prediction indicating whether the code is vulnerable (“yes”), secure (“no”), or unable to answer (“n/a”); and (ii) a summary of the model’s reasoning. The extracted prediction is compared to the ground truth (GT) label to compute the binary accuracy score ($\mathrm{pred}$). To assess reasoning quality, we utilize semantic similarity that evaluates correctness on the basis of embedding similarity and provides a holistic assessment of reasoning quality. The overall scoring metrics contains two components:
\begin{enumerate}
    \item \textbf{Prediction Correctness}: \texttt{gpt-4o-mini} is used to judge whether the LLM’s prediction ($\mathrm{pred}$)  aligns with the ground truth prediction (“yes”/“no”).
    \item \textbf{Reasoning Correctness}: We compute cosine similarity ($\mathrm{cs}$) between the reasoning embeddings of the LLM and the ground truth, using the \texttt{text-embedding-3-small} model. %This enables graded semantic similarity.
\end{enumerate}

The final evaluation metric combines all components into the \textbf{Scoring Metrics ($SM$)} as:
% \begin{equation}
% % \text{SM} = \min\Big(1.0,\;(w_1\times\text{pred}) + (w_2\times\text{cs}) \Big)//
% \text{SM} = \min\Big(1.0,\;(w_1\times\mathrm{pred}) + (w_2\times\mathrm{cs}) \Big)
% \label{eq:fcs}
% \end{equation}
\begin{equation}
\mathit{SM} = \min\Big(1.0,\;(\mathit{w}_1 \times \mathrm{pred}) + (\mathit{w}_2 \times \mathrm{cs}) \Big)
\label{eq:fcs}
\end{equation}
We adopt a weighted scheme for prediction accuracy and reasoning $(w_1 = 0.6,\; w_2 = 0.4)$ to prioritize vulnerability identification over reasoning, a view suited to security-critical contexts where correctly identifying a flaw is the primary objective. This choice is consistent with certain upcoming frameworks, like SIEV~\cite{abbasloo2025measuring}, which also assigns higher weightage to the final outcome ($\lambda = 0.7$) and treats reasoning as a supporting factor. Employing these weights enable us to move beyond static correctness to ensure that model outputs are reliable and not merely the result of surface-level pattern matching. In our case, reasoning is also assigned a significant weight (0.4) because a correct prediction based on flawed logic provides less actionable value and may even dangerously mislead security analysts.

To demonstrate the utility of our metric, we consider a code snippet with CVE-2023-23143 (CWE-120, Buffer Copy without Checking Size of Input) in GPAC's H.264/AVC slice parser. As per ground truth the code is \textit{vulnerable} due to improper bounds checking when bitstream-derived values (such as \texttt{pps\_id} and \texttt{si->pps->sps\_id}) are used directly as array indices (for \texttt{avc->pps} and \texttt{avc->sps}). This can lead to out-of-bounds writes when these indices exceed valid limits (e.g., \texttt{pps\_id = 255}, \texttt{sps\_id = 32}).

We evaluate this snippet using four models employing the CoT prompt with real-world context. GPT-4 incorrectly predicts \textit{secure}. This yields $pred = 0$ and $cs = 0.37$, resulting in $SM = 0 + (0.4 \times 0.37) \approx 0.15$. In this case the reasoning contradicts the ground truth, as the model overlooks the absence of bounds checking and asserts that sufficient input validation is present.

Llama-3-8B correctly predicts \textit{vulnerable} but instead of identifying the specific issue of unchecked index bounds, it focuses on generic buffer overflows, uninitialized variables, and unrelated code regions. Despite $pred = 1$, semantic similarity remains low ($cs = 0.35$), yielding $SM = 0.6 + (0.4 \times 0.35) \approx 0.74$. This indicates correct prediction but poor reasoning.

Qwen2.5-Coder-14B also predicts \textit{vulnerable} and explicitly identifies improper input validation during slice parsing leading to out-of-bounds array accesses, directly corresponding to the misuse of \texttt{pps\_id} and \texttt{sps\_id}. This results in $pred = 1$ and high semantic similarity ($cs = 0.58$), yielding $SM = 0.6 + (0.4 \times 0.58) \approx 0.83$. Gemini also correctly identifies the vulnerability and explicitly points to unchecked usage of \texttt{pps\_id} and \texttt{sps\_id}, achieving higher semantic similarity ($cs = 0.73$) and a corresponding $SM \approx 0.89$.

A binary accuracy metric in this case would rank Llama-3-8B, Qwen2.5-Coder-14B, and Gemini outputs as equally correct, whereas our scoring metric provides more depth and utility by individually assessing both prediction correctness and reasoning quality.

%Compared with prior LLM evaluation frameworks for software security, QuiLL provides three key advantages: (1) integration of real-world vulnerability–patch knowledge through retrieval-augmented context, (2) systematic evaluation of structured prompting strategies under both zero-shot and few-shot settings, and (3) a holistic scoring metric that jointly evaluates prediction correctness and reasoning quality.
%{template used}
 % We formed a vector store by extracting data from a NVD database which contained (info about the database like total samples, columns, how many different CWEs, for how many languages, etc).
 % We queried the database for samples belonging to the following CWEs and written in C/C++ and python. For each sample that was extracted, a template string was populated with the fetched data. The template string was:
% This code is part of {row[6]} project, written in {row[7]} and has {row[0]}, which means {row[1]}. In the following code {row[2]}.
% Vulnerable Code:
% % {row[3]}
% Fixed Code:
% % {row[4]}
% This was the commit message for this fix:
%
%Prompt

\section{Experiment: Vulnerability Detection Using QuiLL}

Here we describe our experiment to benchmark LLMs for vulnerability detection in the wild. The goal of this exercise is to demonstrate the workings of our evaluation framework and highlight its utility and strengths in real-world vulnerabilty detection. For this purpose we have chosen a set of popular representative LLMs and a real-world vulnerability dataset from the research literature. 

%This section discusses the LLMs and dataset used in our evaluation framework and highlights the strength when tested under practical, realistic conditions. We designed an experiment that explore the capability of LLMs to detect and reason about real-world software vulnerabilities. The experiment evaluates how effectively the models can predict and explain about real-world vulnerabilities when evaluated with diverse and previously unseen code examples, thus moving beyond handcrafted or synthetic benchmarks.

\subsection{Large Language Models (LLMs)}
\label{sec:llms}
Our choice of LLMs comprises both open-source and proprietary API based systems and models optimized for local deployment. These include GPT-4, Gemini-1.5-Flash-002, Qwen2.5-Coder-14B, Meta-LLaMA-3-8B, and Phi-4. Table~\ref{tab:model} summarizes the technical specifications of the evaluated models, including parameter count, context window size, and training scale.

\begin{table}[t]
\centering
\resizebox{\columnwidth}{!}{%
\begin{tabular}{@{}lccccc@{}}
\toprule
\textbf{Model} & \textbf{Developer} & \textbf{Parameters} & \textbf{Context Window} & \textbf{Training Scale} & \textbf{Access} \\ \midrule
GPT-4 & OpenAI & -- & 128K & Undisclosed & API \\
Gemini-1.5-Flash-002 & Google & -- & 1M & Multimodal & API \\
Qwen2.5-Coder-14B & Alibaba & 14B & 128K & 5.5T Code Tokens & Local \\
LLaMA-3-8B & Meta & 8B & 8K & 15T Tokens & Local \\
Phi-4-Mini & Microsoft & 3.8B & 128K & Synthetic Data & Local \\ \bottomrule
\end{tabular}%
}
\caption{Technical specifications of the evaluated LLMs.}
\label{tab:model}
\end{table}

Gemini-1.5-Flash-002 is selected as a representative closed-source API model known for its high speed performance and context window size of 1 million tokens, which enables analysis of large codebases. This model is also trained on large scale multimodal and web-based datasets, thereby supporting contextual reasoning.

GPT-4 is included as a strong proprietary baseline due to its consistently superior performance across reasoning and code generation benchmarks. The GPT-4 technical report documents improvements of approximately 16 percentage points on the Massive Multitask Language Understanding (MMLU) benchmark i.e., 70\% to 86.4\%, and achieves 67.0\% on HumanEval (compared to 48.1\% for GPT-3), highlighting significant gains in reasoning and code comprehension capabilities \cite{10}.

% In prior work, Ullah et al. \cite{1} evaluated GPT-4; however their study lacked evaluation under realistic scenarios. We include GPT-4 in our study due to its consistently higher performance across diverse benchmarks.  The GPT-4 technical report reports improvements of approximately 16 percentage points on MMLU i.e., 70\% to 86.4\%, and achieves 67.0\% on HumanEval compared to 48.1\% for GPT-3, highlighting significant gains in reasoning and code understanding capabilities \cite{10}. In our evaluation, GPT-4 achieves the highest CP–CR performance (32.8\% in FS settings), confirming its strong ability to identify subtle vulnerability patterns. However, despite being the top performing model, its overall effectiveness remains limited, highlighting the inherent difficulty of the task. Its training on large-scale multimodal dataset further enhances generalization across code datasets. 

We also evaluate a set of open-source LLMs, including Qwen2.5-Coder-14B, Meta-LLaMA-3-8B, and Phi-4, that are optimized for resource efficiency and local deployment. Prior studies indicate that smaller language models, when properly optimized, retain competitive reasoning performance, enabling efficient performance without substantial degradation in reasoning capability~\cite{srivastava2025towards, thelwall2026can}. Our choice of lightweight variants is also influenced by GPU constraints of 12 GB. 

Qwen2.5-Coder-14B is a specialized open-source code-based LLM designed for programming and software analysis tasks. This model is trained extensively on code repositories and technical data, making it ideally suited for processing code.

Meta-LLaMA-3-8B is another lightweight, open-source model optimized for efficiency and deployment on constrained hardware. Phi-4 is a compact reasoning-based model trained on curated and synthetic datasets to enhance performance on limited computational resources.

We believe our choice of models provides insight into how different LLMs, architectures, and parameters impact the identification and reasoning of software vulnerabilities and opens up interesting new directions. In future work we intend to extend our analysis to a larger set of LLMs, including cutting edge and advanced models, for a comprehensive benchmarking exercise.

\subsection{SecLLMHolmes Dataset}

For the purposes of this evaluation, we employ a well-known dataset from the research literature: the SecLLHolmes dataset, curated by Ullah et al.\cite{1}, is a real-world vulnerability dataset containing both vulnerable and patched code snippets. It comprises publicly disclosed vulnerabilities from the Common Vulnerabilities and Exposures (CVE) repository and vulnerabilities reported in the Top CWE categories identified by the MITRE CWE list~\cite{13}.

The original SecLLMHolmes dataset contains 228 code scenarios including 48 hand-crafted ones, 30 real-world instances, and 150 augmented code scenarios. Considering our emphasis on realistic real-world vulnerability detection, we focus specifically on the 30 real-world scenarios subset. This amounts to $\approx 6,700$ lines of truncated code (vulnerable versions only), roughly doubling when including the patched versions ($\approx 13,300$ lines).  This data comprises a total of 15 CVEs belonging to the MITRE Top 25 Most Dangerous Software Weaknesses (2024)~\cite{13}. These snippets belong to four open-source projects (Linux, Libtiff, Gpac and Pjsip) and these vulnerabilities represent widely observed security flaws in practical software systems. The detailed distribution of the selected CVEs is presented in Table~\ref{tab:grouped_cves}.

\begin{table}[t!]
\small
\centering
\renewcommand{\arraystretch}{1.1}
\begin{tabular}{p{1cm}p{2.1cm}p{3.7cm}}
\toprule
\textbf{Rank} & \textbf{CWE ID} & \textbf{Project and CVE ID} \\
\midrule

\multirow{2}{*}{1}
 & \textbf{CWE-787} & 
    gpac: CVE-2023-1452, CVE-2023-23143; \newline
    libtiff: CVE-2023-26966; \newline
    linux: CVE-2023-45863, CVE-2023-45871; \newline
    pjsip: CVE-2023-27585 \\
    
\midrule
4 
 & \textbf{CWE-416} & 
    linux: CVE-2023-40283 \\
    
\midrule
\multirow{1}{*}{12}
 & \textbf{CWE-476} & 
    gpac: CVE-2023-3012; \newline
    libtiff: CVE-2023-2908, CVE-2023-3316; \newline
    linux: CVE-2023-42754 \\
    
\midrule
\multirow{1}{*}{14}
 & \textbf{CWE-190} & 
    gpac: CVE-2023-23144; \newline
    libtiff: CVE-2023-40745, CVE-2023-41175; \newline
    linux: CVE-2023-42753 \\
    
\bottomrule
\end{tabular}
\captionsetup{font=small}
\caption{CVEs grouped by MITRE Ranking and CWE ID. CWE Types: \textbf{CWE-787} = Out-of-Bounds Write, \textbf{CWE-476} = NULL Pointer Dereference, \textbf{CWE-190} = Integer Overflow, \textbf{CWE-416} = Use After-Free.}
\label{tab:grouped_cves}
\end{table}

\subsection{Experiment Modes}
We conduct our experiment in two settings:  
\begin{itemize}
    \item \textbf{Zero-Shot (ZS):} This setting evaluates the intrinsic behavior of LLMs. Each model is asked to identify whether the code is vulnerable or not. If a vulnerability is detected, the model must identify its location and reason about it.
    \item In the \textbf{Few-Shot (FS):} In this setting, we employ knowledge augmentation via retrieval-augmented generation (RAG) to enhance LLMs reasoning capability.
\end{itemize}

The SecLLMHolmes real-world subset comprises 15 top-ranked CVEs from MITRE's database for both 2023 and 2024, each consisting of both a vulnerable and a patched version, totaling 30 code snippets. For each snippet, we query our five representative LLMs under both ZS and FS settings. We undertake two independent queries per case to account for stochastic variation. This results in 240 runs per model (15 CVEs × 2 samples per CVE × 2 settings × 2 queries).

Conducting dual queries per sample helps capture behavioral variation and assess whether contextual augmentation improves the model’s ability to identify vulnerabilities. For instance, a model might fail to detect a buffer overflow in the ZS setting but successfully identify it in the FS setting when given a similar vulnerable–patch pair.

\section{Results and Analysis}

This section systematically analyzes the performance of LLMs across multiple dimensions of vulnerability detection. We first examine overall predictive accuracy and reasoning reliability under ZS and FS settings to assess deployment readiness. Next, we investigate reasoning failures and their impact on prediction quality, followed by an analysis of overlap in detected vulnerabilities across models to understand generalization patterns. We then evaluate the effectiveness of different prompting strategies in improving reasoning consistency and accuracy. Finally, we assess the proposed scoring metric, which jointly captures prediction correctness and reasoning quality. We summarize below key insights from our results:

\subsection{\textbf{LLMs are Not Yet Ready for Real-World Deployment}}
Fig.~\ref{fig:merged-bar} summarizes model performance across ZS and FS settings. While FS prompting substantially improves both predictive accuracy and reasoning quality across all evaluated models. However despite these gains, LLMs still exhibit significant inconsistency and reasoning errors in real-world vulnerability detection. Overall reliability remains insufficient for real-world deployment.

The proportion of correct predictions with correct reasoning (CP-CR) is consistently higher in FS than in ZS settings but remains below 40\% for all models. 

For instance, GPT-4 achieves the highest CP-CR percentage at 32.8\% in FS, compared to only 22\% in ZS. Similar improvements are observed for Qwen2.5-Coder and Phi-4, which reach 31.3\% and 25.0\% CP-CR in FS, but drop to 22.0\% and 18.8\% in ZS, respectively. This indicates a notable dependence on contextual examples for reliable reasoning Despite improvement, the result indicate that even the strongest models frequently misclassify or provide flawed justifications.

A key factor in this regard may be model hallucination, whereby a model generates fluent output unsupported by facts or inputs \cite{alansari2026large}. LLMs are next word predictors, they cannot reliably backtrack, and early mistakes in reasoning can propagate to the final answer \cite{song2026large}. Shen et al. also demonstrate that LLMs may arrive at correct answers while relying on reasoning paths that are inconsistent or irrelevant, highlighting that such explanations can be post hoc rationalizations rather than causally grounded reasoning \cite{shen2025faithcot}.
 
 For the case of vulnerability detection, this limitation is not acceptable as this task requires precise semantic understanding (control/data flow) of code and reasoning capabilities that LLMs thus far approximate imperfectly. 
 
The performance of LLMs also varies widely due to scale and design. As parameter count increases, model capabilities improve, following predictable mathematical patterns referred to as \textit{scaling laws} \cite{kaplan2020scaling}. The parameter count of proprietary models like GPT-4 and Gemini has not been officially disclosed. However, an estimated count demonstrates that GPT-4 has parameters in the trillions \cite{wiki_gpt4_2024} whereas Qwen2.5-Coder has 14 billion parameters \cite{hui2024qwen2}, which we believe explains the observed difference in their performance.
 
 A key observation is that contextual augmentation improves performance but does not fundamentally resolve reasoning quality. In ZS settings models often fail to predict and reason correctly, particularly on vulnerable code samples. It does not yield the level of reliability required for vulnerability detection in real-world. Models frequently produce correct answer yet incorrect reasoning (CP–ICR) or fail to precisely localize vulnerabilities.  In safety critical areas, such response can lead to false confidence and unsafe decisions.

Our results indicate that, although LLMs demonstrate promise in vulnerability detection, their current levels of reasoning consistency and interpretability falls far short of the requirements for unsupervised deployment in real-world security pipelines.

\begin{figure}[t!]
\centering

\begin{tikzpicture}
\begin{axis}[
    xbar stacked,
    bar width=6pt,
    width=0.43\textwidth,
    height=7.2cm,
    xmin=0, xmax=120,
    xlabel={Number of Samples},
    ytick=data,
    yticklabels={
        Phi-4 (FS),
        Phi-4 (ZS),
        LLaMA-3 (FS),
        LLaMA-3 (ZS),
        Qwen2.5-Coder (FS),
        Qwen2.5-Coder (ZS),
        GPT-4 (FS),
        GPT-4 (ZS),
        Gemini (FS),
        Gemini (ZS)
    },
    enlarge y limits=0.05,
    axis x line*=bottom,
    axis y line*=left,
    tick label style={font=\scriptsize},
    label style={font=\scriptsize},
    xtick style={draw=none},
    legend style={font=\scriptsize, at={(0.5,1.05)}, anchor=south, legend columns=3}
]
% Green (CP-CR)
\addplot+[xbar, fill={rgb,255:red,102;green,194;blue,165}] coordinates {
    (25,0) (15,1)
    (27,2) (23,3) %27 LLaMA
    (21,4) (19,5)%21 Qwen2.5-Coder
    (38,6) (22,7)%38 GPT-4  ,22 zs gpt-4
    (23,8) (18,9)  %23 Gemini
};
% Yellow (CP-ICR)
\addplot+[xbar, fill={rgb,255:red,255;green,217;blue,102}] coordinates {
    (22,0) (14,1)
    (20,2) (24,3) %20 LLaMA
    (13,4) (10,5)%13 Qwen2.5-Coder
    (20,6) (23,7)%20 GPT-4, 23 zs gpt-4
    (36,8) (30,9) %36 Gemini
};
% Red (ICP-ICR)
\addplot+[xbar, fill={rgb,255:red,250;green,128;blue,114}] coordinates {
    (33,0) (51,1)
    (43,2) (52,3) %43 LLaMA
    (33,4) (34,5) %33 Qwen2.5-Coder
    (58,6) (55,7) %58 GPT-4
    (56,8) (65,9) %56 Gemini
};
% \legend{CP-CR, CP-ICR, ICP-ICR}
\end{axis}
\end{tikzpicture}
\caption{
\textbf{Model Performance on Vulnerability Detection and Reasoning.}
Each model is evaluated across two experimental settings, Zero-Shot (ZS) and Few-Shot (FS), each represented by a horizontal bar comprising 120 total evaluation instances.
Bar segments indicate the distribution of outcomes:
\textcolor[rgb]{0.40,0.76,0.65}{CP-CR (Correct Prediction \& Correct Reason)}, 
\textcolor[rgb]{1.00,0.85,0.40}{CP-ICR (Correct Prediction \& Incorrect Reason)},  \textcolor[rgb]{0.98,0.50,0.45}{ICP-ICR (Incorrect Prediction \& Incorrect Reason)}.
}
\label{fig:merged-bar}
\end{figure}

% \subsection{Incorrect Reasoning is a Significant Problem}
% \subsection{Persistent reasoning flaws remain a critical barrier to reliable detection}
\subsection{\textbf{Incorrect Reasoning is a Critical Bottleneck}}
The proportion of incorrect prediction accompanied by incorrect reasoning (ICP-ICR) increases significantly in the absence of contextual information, underscoring a critical weakness in current LLM reasoning. Phi-4 is particularly sensitive, with ICP-ICR increasing from 41.3\% in FS to 63.8\% in ZS setting, a degradation of 22.5\%. Gemini follows a similar trend, increase from 48.7\%  to 57.5\% without contextual augmentation. 

When focusing on correct predictions, the reasoning quality remains inconsistent. Gemini achieves a CP-CR of only 19.2\% in FS setting and has a high CP-ICR rate of 31.3\%, suggesting that even when the model’s output is correct, its reasoning is frequently incomplete or logically flawed. 

In contrast, LLaMA-3 shows balanced performance with a CP-CR of 30.0\% and a moderate ICP-ICR of 47.8\% in the FS scenario, indicating relatively stronger reasoning consistency compared to other open-source models. Overall, FS settings reduce ICP-ICR by 4–22\% across models and improve CP-CR by 4–13\%. These results highlight that while FS helps, incorrect reasoning remains a significant barrier to reliable and interpretable LLM-based vulnerability detection. Persistent reasoning failures limit model reliability and highlight the need for structured reasoning mechanisms or manual verification before real-world deployment.

%Reasoning quality remains the primary bottleneck.
Among all the models, closed-source models GPT-4 and Gemini outperforms open-source models particularly when provided with context. The reasoning quality remains the primary bottleneck. LLMs frequently output correct answers through flawed or incomplete reasoning, especially in ZS scenarios. Such behavior is risky in safety-critical applications where reasoning consistency is as important as the output itself. This finding highlights the urgent need for mechanisms that enforce reasoning validation and consistency checks in LLM-based vulnerability analysis.\\

\subsection{\textbf{Detected Vulnerabilities Overlap Significantly}}
\begin{figure*}[t!]
\centering
\includegraphics[width=0.9\textwidth]{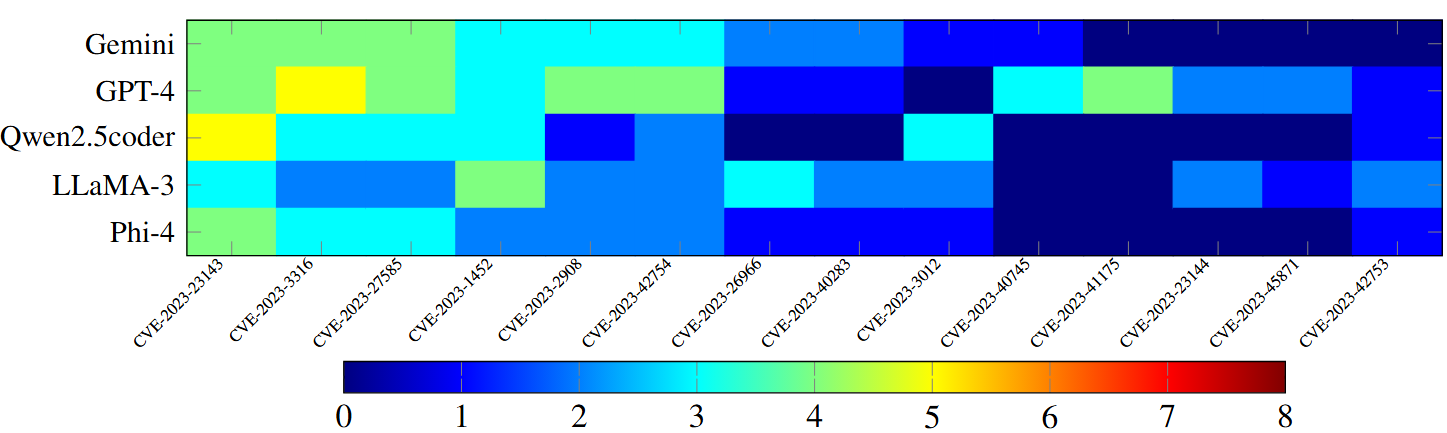}
\captionsetup{position=above, font=small} % caption above
\caption{\textbf{Correct Predictions with Correct Reasoning — Heatmap (Models × CVEs)}}
\label{fig:CVE2}
\end{figure*}
% ~\ref{fig:cve-barplot}
Fig.~\ref{fig:CVE2} presents a heatmap of correct predictions with correct reasoning across individual CVEs for all evaluated models. The visualization reveals a strong overlap in the vulnerabilities that multiple models are able to detect and reason about correctly. Certain CVEs, such as \texttt{CVE-2023-23143}, \texttt{CVE-2023-3316}, and \texttt{CVE-2023-27585}, are consistently detected with correct reasoning across nearly all models. 

In contrast, several CVEs including \texttt{CVE-2023-41175}, \texttt{CVE-2023-23144}, and \texttt{CVE-2023-45871} fails to produce any correct reasoning from any model. Among the evaluated models, GPT-4 demonstrates the broadest coverage of CVEs with correct reasoning, whereas Qwen2.5-Coder and Gemini show fewer fully incorrect cases, indicating a stable reasoning pattern. Qwen2.5-Coder, LLaMA-3, and Phi-4 perform well only on a smaller subset of the same CVEs, further reinforcing the presence of a shared knowledge boundary across architectures.

%This consistent failure indicates that LLMs struggle with less common or more complex vulnerability types, particularly vulnerability patterns not seen during training.

These results suggests that these vulnerabilities exhibit structural or semantic patterns that align well with the pretrained knowledge of LLMs. This result is consistent with documented findings in the literature \cite{li2025everything}, indicating that LLMs struggle with rare vulnerabilities that are unseen during training. LLMs tend to generalize along familiar vulnerability patterns and struggle with rare or complex cases, which emphasizes the urgent need for more diverse training data \cite{li2025everything}.

%heatmap of CVEs
%We observe that model performance varies significantly across different CVEs. Some vulnerabilities (e.g., those with familiar CWE patterns) are consistently detected, while others receive no correct answers. This suggests that LLMs generalize better to patterns likely seen during pretraining but struggle with novel or unseen vulnerabilities. Although GPT-4 leads in total correct responses, Qwen2.5-Coder and Gemini show fewer fully incorrect cases, indicating a stable reasoning pattern.

\begin{figure}[b!]
\centering
\begin{tikzpicture}
\begin{axis}[
    width=0.46\textwidth,
    height=0.38\textwidth,
    xlabel={\scriptsize Prompt},
    ylabel={\scriptsize Correct Prediction \& Correct Reasoning},
    ymin=0, ymax=30,
    ytick={0,5,10,15,20,25,30},
    xtick=data,
    xticklabel style={font=\tiny},
    yticklabel style={font=\tiny},
    legend style={
        at={(0.97,0.97)},
        anchor=north east,
        legend columns=1,
        font=\tiny,
        inner sep=2pt,
        row sep=0pt
    },
    grid=both,
    grid style={line width=.1pt, draw=gray!30},
    major grid style={line width=.2pt, draw=gray!50},
    tick style={black},
    axis line style={black},
    symbolic x coords={P-S,P-CoT,P-Decomp,P-P\&S},
    enlargelimits=0.05,
    smooth
]
% Gemini
\addplot+[mark=*, thick, green!60!black] coordinates {
    (P-S,8) (P-CoT,7) (P-Decomp,6) (P-P\&S,6)
};
\addlegendentry{Gemini}
% GPT-4
\addplot+[mark=square*, thick, blue] coordinates {
    (P-S,9) (P-CoT,9) (P-Decomp,9) (P-P\&S,11)
};
\addlegendentry{GPT-4}
% Qwen2.5-Coder
\addplot+[mark=triangle*, thick, orange] coordinates {
    (P-S,6) (P-CoT,5) (P-Decomp,6) (P-P\&S,4)
};
\addlegendentry{Qwen2.5-Coder}
% LLaMA-3
\addplot+[mark=o, thick, yellow!80!black] coordinates {
    (P-S,5) (P-CoT,8) (P-Decomp,6) (P-P\&S,8)
};
\addlegendentry{LLaMA-3}
% Phi-4
\addplot+[mark=diamond*, thick, lime!70!black] coordinates {
    (P-S,4) (P-CoT,5) (P-Decomp,6) (P-P\&S,5)
};
\addlegendentry{Phi-4}
\end{axis}
\end{tikzpicture}
\captionsetup{font=small}
\caption{
\textbf{Correct Predictions \& Correct Reasoning by Prompt.}
Each line shows model performance under four prompts---P-S (Standard), P-CoT (Chain-of-Thought), P-Decomp (Decomposition), and P-P\&S (Plan-and-Solve) for vulnerability detection in the wild.}
\label{fig:correct-P-line-2}
\end{figure}

\begin{figure}[b!]
    \centering
    \captionsetup{font=small}
\includegraphics[width=0.48\textwidth]{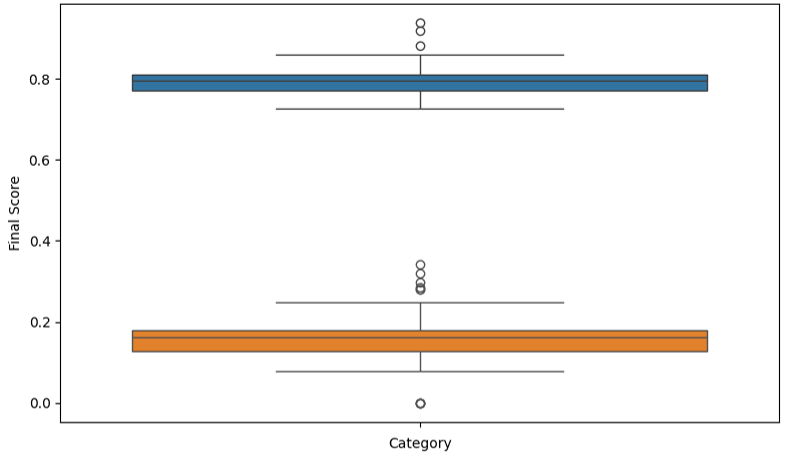}
\caption{Distribution of Scoring Metrics that jointly evaluates prediction and its reasoning correctness}
    \label{fig:SM}
\end{figure}

% \subsection{Best Prompt}
\subsection{\textbf{Prompting Strategies Significantly Enhance Detection and Consistency}}
Fig.~\ref{fig:correct-P-line-2} compares LLMs performance across four prompting strategies: Standard Prompt (P-S), Chain-of-Thought Prompt (P-CoT), Decomposition Prompt (P-Decomp), and Plan-and-Solve Prompt (P-P\&S). P-P\&S prompt achieves the highest overall performance, particularly for GPT-4 and LLaMA-3. This prompt explicitly encourages models to outline a reasoning plan before reaching conclusions, leading to more coherent and accurate vulnerability assessments.

While P-P\&S yields the strongest performance, the P-Decomp prompt demonstrates the most consistent results across all evaluated models. Unlike other strategies that occasionally degrade performance for smaller models, P-Decomp maintains steady accuracy by guiding models through a structured step-by-step reasoning process that mirrors the analytical workflow of human security experts: understanding the code, localizing potential flaws, and justifying the vulnerability with reasoning. GPT-4 demonstrates the best performance overall with P-P\&S. 

The P-CoT prompt improves reasoning consistency across models, confirming that encouraging step-wise reasoning enhances model interpretability.

%prompt design plays a decisive role
Structured reasoning prompts (P-CoT, P-Decomp, P-P\&S) improve correct predictions by roughly 6–10 \% on average, and substantially more for large models such as for GPT-4. These prompts yields consistent performance across all models, making them strong general-purpose strategies. This aligns with prior evidence that modular reasoning approaches are more effective for complex analytical tasks~\cite{71,72}. %GPT-4 achieves its best results with Plan-and-Solve prompt, while P-Decomp offers the most consistent performance across models. This highlights and confirms the advantage of explicit structured reasoning and step-by-step planning. It can partially compensate for limited security-domain expertise within the models. 

% \subsection{Benefit of Evaluation Metrics}
\subsection{\textbf{Our Scoring Metric Successfully Reflects Prediction Accuracy and Reasoning Quality}}

Fig.~\ref{fig:SM} presents the distribution of our novel scoring metric, which jointly assesses both prediction accuracy and its reasoning quality. We observe a separation between categories: results with correct prediction and correct reasoning (CP-CR) cluster score above 0.75, while those with incorrect predictions and incorrect reasoning (ICP-ICR) remain below 0.2. This clear distinction highlights the effectiveness of our metric in capturing both the correctness and reasoning quality of models, providing a more informative and comprehensive evaluation than binary accuracy alone.

Based upon our evaluation, the samples achieving fully correct predictions with high confidence reasoning (\textit{acc}=1, \textit{cs}$\geq$0.55) attain scores in the range of \textbf{0.82-0.95}, confirming that the metric rewards both accurate outputs and correct explanations. Partially correct samples (\textit{acc}=1, \textit{cs}$<$0.55), where predictions are accurate but reasoning confidence falls below threshold score between \textbf{0.67-0.82}, reflecting a penalty for insufficient reasoning quality despite correct outcomes.

Conversely, partially incorrect samples (\textit{acc}=0, \textit{cs}$\geq$0.40) score in the \textbf{0.16-0.33} range, indicating that partial reasoning similarity provides only marginal credit in the case of incorrect prediction. Finally, fully incorrect samples (\textit{acc}=0, \textit{cs}$<$0.40) receive the lowest scores of \textbf{0.08-0.16}, capturing the compounded failure of both incorrect prediction and poor reasoning.

%Evaluation Metrics
The proposed evaluation scoring metric demonstrates its effectiveness of combined detection accuracy and reasoning quality spanning from $<$0.2 for complete incorrect answers to $>$0.82 for fully correct responses. The clear separation between high and low scoring samples validates the metric’s robustness for integrated reasoning and prediction assessment over accuracy alone.

\subsection{\textbf{Contextual Aiding (RAG) Improves Performance}}
%Detection in the wild setting i.e., in the wild is substantially harder than in synthetic settings. The results shows that the model, although capable of reasoning about code semantics, are not ready to be deployed in the wild. The results are in consistent with the findings of \cite{1}.
%contextual augmentation (Few-Shot setting)

Our findings confirm that contextual augmentation (FS) significantly enhances accuracy across all models as compared to ZS settings. Providing semantically similar vulnerable and patch pairs improves model understanding of structural code flaws, reducing reasoning errors by up to 22\%. This demonstrates that retrieval-based contextual augmentation is an effective mechanism for compensating for lack of domain-specific knowledge in LLMs.

Whereas this is a marked improvement, we observe that, even under FS scenario, CP–CR remains below 40\% across models, underscoring our earlier claim that LLMs are not yet ready for real-world deployments.

\subsection{\textbf{Large Closed-Source Models Perform Better}}

Our results show that large closed-source models like GPT-4 and Gemini significantly outperform open-source counterparts. However, performance drops sharply when detecting vulnerabilities (as opposed to secure code), especially in the ZS setting. Even GPT-4, which correctly identifies secure samples in FS mode, evidences limited success with vulnerable samples.

These results emphasize yet again the inherent difficulty vulnerability detection in the wild. None of the models in our experiment demonstrate the consistency, interpretability, or reasoning capacity required thus far for unsupervised real-world deployment.

%In summary, the findings highlight that while Future efforts should focus on fine-tuning LLMs with vulnerability-specific, reasoning-driven data, and leveraging structured prompts. With such advancements, LLMs hold strong potential to evolve from supportive vulnerability analysis tools to fully dependable tool in automated cybersecurity workflows.
\section{Conclusion \& Future Work}

In this paper we present QuiLL, the first evaluation framework specifically designed to benchmark LLMs for software vulnerability detection and reasoning in the \textit{wild}. QuiLL provides three key advantages over prior LLM evaluation frameworks: (1) integration of real-world vulnerability–patch knowledge through retrieval-augmented context, (2) systematic evaluation of structured prompting strategies under both zero-shot and few-shot settings, and (3) a holistic scoring metric that jointly evaluates prediction correctness and reasoning quality.

%Our framework integrates carefully designed prompting strategies, real-world CVEs, and in-context examples which are selected on the basis of semantically similarity, enabling a practical and reproducible evaluation setting.

To showcase the strengths of our framework, we evaluate a set of prominent LLMs, including GPT-4, Gemini-1.5-flash, Qwen2.5-Coder, Phi-4 and LLaMA-3 on a well-known dataset of real-world vulnerabilities. Our results confirm and expand on several insights in the research literature and highlight the effectiveness of our framework in capturing the unique nuances and complexities of vulnerability detection in the wild. These particularly include the role of different prompting strategies, knowledge augmentation, and the need for diverse training data.

%Our results signifies that closed-source models outperforms open-source models when guided by structured, role-based, or in-context prompts. A key insight from our framework is that Plan-and-Solve strategy leads to performance improvement in terms of vulnerability detection and its reasoning across closed source models verified by our assessment scoring metrics.

%Open-source models, while showing more limited overall capability, performed relatively better when prompted with simple and direct instructions. Among them, Alibaba's model Qwen2.5-Coder frequently outputs unstable response despite being trained on code, whereas LLaMA-3, showed more stable performance.

However, our results also confirm marked limitations of LLMs in real-world settings. Results across a range of CVEs indicate that LLMs currently do not exhibit the accuracy and reasoning capability to generalize well across a wide range of software vulnerabilities. This wide gap between model capability and real-world requirements indicates considerable opporunities for the research community.

In future work, we aim to extend QuiLL with agentic AI workflows capable of collecting contextual information from multiple sources. This will aid in developing adaptive and robust vulnerability detection and reasoning pipelines, moving closer to practical integration of LLMs into software security workflows.

% Recent surveys highlight that current LLM-based vulnerability detection methods face several critical limitations rooted in the research literature. Sheng et al. observe that existing LLM approaches struggle with cross-language vulnerability detection, multimodal integration, and repository-level analysis, and that existing benchmarks and datasets often lack the scale and contextual richness required for robust evaluation, while also underscoring that model interpretability remains an open challenge. Similarly, Taghavi Far & Feyzi point out that the performance of LLMs in vulnerability detection is constrained by the availability of high-quality labeled datasets and that biases inherited from training data can affect reliability, especially in security-critical applications.

\bibliographystyle{IEEEtran}
\bibliography{IEEEabrv,ref}

\begin{thebibliography}{10}
\providecommand{\url}[1]{#1}
\csname url@samestyle\endcsname
\providecommand{\newblock}{\relax}
\providecommand{\bibinfo}[2]{#2}
\providecommand{\BIBentrySTDinterwordspacing}{\spaceskip=0pt\relax}
\providecommand{\BIBentryALTinterwordstretchfactor}{4}
\providecommand{\BIBentryALTinterwordspacing}{\spaceskip=\fontdimen2\font plus
\BIBentryALTinterwordstretchfactor\fontdimen3\font minus
  \fontdimen4\font\relax}
\providecommand{\BIBforeignlanguage}[2]{{%
\expandafter\ifx\csname l@#1\endcsname\relax
\typeout{** WARNING: IEEEtran.bst: No hyphenation pattern has been}%
\typeout{** loaded for the language `#1'. Using the pattern for}%
\typeout{** the default language instead.}%
\else
\language=\csname l@#1\endcsname
\fi
#2}}
\providecommand{\BIBdecl}{\relax}
\BIBdecl

\bibitem{nist_nvd_statistics}
{National Institute of Standards and Technology}, ``Nvd – search \&
  statistics,''
  \url{https://nvd.nist.gov/vuln/search#/nvd/home?resultType=statistics}, 2025,
  accessed: 2025-10-13.

\bibitem{ibm_2024}
D.~Bonderud, ``Cost of a data breach 2024: Financial industry,''
  \url{https://www.ibm.com/think/insights/cost-of-a-data-breach-2024-financial-industry?mhsrc=ibmsearch_a&mhq=cost%20of%20data%20breach%20report%202024},
  IBM, 2024, accessed: 2025-10-14.

\bibitem{63}
\BIBentryALTinterwordspacing
{IBM Corporation}, ``What is log4j? understanding the cybersecurity
  vulnerability,'' \url{https://www.ibm.com/think/topics/log4j}, 2023,
  accessed: 2025-07-19. [Online]. Available:
  \url{https://www.ibm.com/think/topics/log4j}
\BIBentrySTDinterwordspacing

\bibitem{2023_moveit}
\BIBentryALTinterwordspacing
M.~J. Schwartz, ``Known moveit attack victim count reaches 2,618
  organizations,'' 2023, accessed: 2025-10-14. [Online]. Available:
  \url{https://www.bankinfosecurity.com/known-moveit-attack-victim-count-reaches-2618-organizations-a-23640}
\BIBentrySTDinterwordspacing

\bibitem{lipp2022empirical}
S.~Lipp, S.~Banescu, and A.~Pretschner, ``An empirical study on the
  effectiveness of static c code analyzers for vulnerability detection,'' in
  \emph{Proceedings of the 31st ACM SIGSOFT international symposium on software
  testing and analysis}, 2022, pp. 544--555.

\bibitem{19}
Z.~Li, D.~Zou, S.~Xu, X.~Ou, H.~Jin, S.~Wang, Z.~Deng, and Y.~Zhong,
  ``Vuldeepecker: A deep learning-based system for vulnerability detection,''
  \emph{arXiv preprint arXiv:1801.01681}, 2018.

\bibitem{8}
M.~Fu and C.~Tantithamthavorn, ``Linevul: A transformer-based line-level
  vulnerability prediction,'' in \emph{Proceedings of the 19th International
  Conference on Mining Software Repositories}, 2022, pp. 608--620.

\bibitem{34}
A.~Khare, S.~Dutta, Z.~Li, A.~Solko-Breslin, R.~Alur, and M.~Naik,
  ``Understanding the effectiveness of large language models in detecting
  security vulnerabilities,'' \emph{arXiv preprint arXiv:2311.16169}, 2023.

\bibitem{7}
R.~Safdar, M.~U. Ashfaq, and D.~Mateen, ``Deep learning-based framework for
  automated vulnerability detection in android applications,'' in \emph{2023
  20th International Bhurban Conference on Applied Sciences and Technology
  (IBCAST)}.\hskip 1em plus 0.5em minus 0.4em\relax IEEE, 2023, pp. 1--5.

\bibitem{nong2025appatch}
Y.~Nong, H.~Yang, L.~Cheng, H.~Hu, and H.~Cai, ``$\{$APPATCH$\}$: Automated
  adaptive prompting large language models for $\{$Real-World$\}$ software
  vulnerability patching,'' in \emph{34th USENIX Security Symposium (USENIX
  Security 25)}, 2025, pp. 4481--4500.

\bibitem{ding2024vulnerability}
Y.~Ding, Y.~Fu, O.~Ibrahim, C.~Sitawarin, X.~Chen, B.~Alomair, D.~Wagner,
  B.~Ray, and Y.~Chen, ``Vulnerability detection with code language models: How
  far are we?'' \emph{arXiv preprint arXiv:2403.18624}, 2024.

\bibitem{safdar2025data}
R.~Safdar, D.~Mateen, S.~T. Ali, M.~U. Ashfaq, and W.~Hussain, ``Data and
  context matter: Towards generalizing ai-based software vulnerability
  detection,'' \emph{arXiv preprint arXiv:2508.16625}, 2025.

\bibitem{chakraborty2021deep}
S.~Chakraborty, R.~Krishna, Y.~Ding, and B.~Ray, ``Deep learning based
  vulnerability detection: Are we there yet?'' \emph{IEEE Transactions on
  Software Engineering}, vol.~48, no.~9, pp. 3280--3296, 2021.

\bibitem{1}
S.~Ullah, M.~Han, S.~Pujar, H.~Pearce, A.~Coskun, and G.~Stringhini, ``Llms
  cannot reliably identify and reason about security vulnerabilities (yet?): A
  comprehensive evaluation, framework, and benchmarks,'' in \emph{2024 IEEE
  Symposium on Security and Privacy (SP)}.\hskip 1em plus 0.5em minus
  0.4em\relax IEEE, 2024, pp. 862--880.

\bibitem{35}
Z.~Gao, H.~Wang, Y.~Zhou, W.~Zhu, and C.~Zhang, ``How far have we gone in
  vulnerability detection using large language models,'' \emph{arXiv preprint
  arXiv:2311.12420}, 2023.

\bibitem{37}
Y.~Liu, L.~Gao, M.~Yang, Y.~Xie, P.~Chen, X.~Zhang, and W.~Chen,
  ``Vuldetectbench: Evaluating the deep capability of vulnerability detection
  with large language models,'' \emph{arXiv preprint arXiv:2406.07595}, 2024.

\bibitem{38}
A.~Zibaeirad and M.~Vieira, ``Vulnllmeval: A framework for evaluating large
  language models in software vulnerability detection and patching,''
  \emph{arXiv preprint arXiv:2409.10756}, 2024.

\bibitem{9}
Y.~Sun, D.~Wu, Y.~Xue, H.~Liu, W.~Ma, L.~Zhang, Y.~Liu, and Y.~Li, ``Llm4vuln:
  A unified evaluation framework for decoupling and enhancing llms'
  vulnerability reasoning,'' \emph{arXiv preprint arXiv:2401.16185}, 2024.

\bibitem{42}
M.~Chandramohan, Y.~Xue, Z.~Xu, Y.~Liu, C.~Y. Cho, and H.~B.~K. Tan, ``Bingo:
  Cross-architecture cross-os binary search,'' in \emph{Proceedings of the 2016
  24th ACM SIGSOFT International Symposium on Foundations of Software
  Engineering}, 2016, pp. 678--689.

\bibitem{43}
B.~Johnson, Y.~Song, E.~Murphy-Hill, and R.~Bowdidge, ``Why don't software
  developers use static analysis tools to find bugs?'' in \emph{2013 35th
  International Conference on Software Engineering (ICSE)}.\hskip 1em plus
  0.5em minus 0.4em\relax IEEE, 2013, pp. 672--681.

\bibitem{44}
J.~Smith, B.~Johnson, E.~Murphy-Hill, B.~Chu, and H.~R. Lipford, ``Questions
  developers ask while diagnosing potential security vulnerabilities with
  static analysis,'' in \emph{Proceedings of the 2015 10th Joint Meeting on
  Foundations of Software Engineering}, 2015, pp. 248--259.

\bibitem{45}
J.~Newsome and D.~X. Song, ``Dynamic taint analysis for automatic detection,
  analysis, and signaturegeneration of exploits on commodity software.'' in
  \emph{NDSS}, vol.~5.\hskip 1em plus 0.5em minus 0.4em\relax Citeseer, 2005,
  pp. 3--4.

\bibitem{baldoni2018survey}
R.~Baldoni, E.~Coppa, D.~C. D’elia, C.~Demetrescu, and I.~Finocchi, ``A
  survey of symbolic execution techniques,'' \emph{ACM Computing Surveys
  (CSUR)}, vol.~51, no.~3, pp. 1--39, 2018.

\bibitem{kirda2006behavior}
E.~Kirda, C.~Kruegel, G.~Banks, G.~Vigna, and R.~Kemmerer, ``Behavior-based
  spyware detection.'' in \emph{Usenix Security Symposium}, 2006, p. 694.

\bibitem{46}
M.~Monga, R.~Paleari, and E.~Passerini, ``A hybrid analysis framework for
  detecting web application vulnerabilities,'' in \emph{2009 ICSE Workshop on
  Software Engineering for Secure Systems}.\hskip 1em plus 0.5em minus
  0.4em\relax IEEE, 2009, pp. 25--32.

\bibitem{shields2023hybrid}
P.~Shields, ``Hybrid testing: Combining static analysis and directed fuzzing,''
  Ph.D. dissertation, Massachusetts Institute of Technology, 2023.

\bibitem{31}
F.~Yamaguchi, N.~Golde, D.~Arp, and K.~Rieck, ``Modeling and discovering
  vulnerabilities with code property graphs,'' in \emph{2014 IEEE symposium on
  security and privacy}.\hskip 1em plus 0.5em minus 0.4em\relax IEEE, 2014, pp.
  590--604.

\bibitem{sheng2025llms}
Z.~Sheng, Z.~Chen, S.~Gu, H.~Huang, G.~Gu, and J.~Huang, ``Llms in software
  security: A survey of vulnerability detection techniques and insights,''
  \emph{ACM Computing Surveys}, vol.~58, no.~5, pp. 1--35, 2025.

\bibitem{25}
H.~Zheng, L.~Shen, Y.~Luo, T.~Liu, J.~Shen, and D.~Tao, ``Decomposed prompt
  decision transformer for efficient unseen task generalization,''
  \emph{Advances in Neural Information Processing Systems}, vol.~37, pp.
  122\,984--123\,006, 2024.

\bibitem{64}
X.~Du, G.~Zheng, K.~Wang, Y.~Zou, Y.~Wang, W.~Deng, J.~Feng, M.~Liu, B.~Chen,
  X.~Peng \emph{et~al.}, ``Vul-rag: Enhancing llm-based vulnerability detection
  via knowledge-level rag,'' \emph{arXiv preprint arXiv:2406.11147}, 2024.

\bibitem{14}
G.~Bhandari, A.~Naseer, and L.~Moonen, ``Cvefixes: automated collection of
  vulnerabilities and their fixes from open-source software,'' in
  \emph{Proceedings of the 17th International Conference on Predictive Models
  and Data Analytics in Software Engineering}, 2021, pp. 30--39.

\bibitem{qdrant2026}
{Qdrant Team}, ``Qdrant: Open-source vector database for ai applications,''
  \url{https://qdrant.tech/}, 2026, accessed: 2026-03-18.

\bibitem{16}
\BIBentryALTinterwordspacing
{OpenAI}, ``Openai embeddings documentation,'' 2024, accessed: 2025-04-24.
  [Online]. Available: \url{https://platform.openai.com/docs/guides/embeddings}
\BIBentrySTDinterwordspacing

\bibitem{66}
{LangChain}, ``Openai embeddings integration,''
  \url{https://js.langchain.com/docs/integrations/text_embedding/openai}, 2024,
  accessed: 2025-07-25.

\bibitem{67}
{LlamaIndex}, ``Using openai embeddings in llamaindex,''
  \url{https://docs.llamaindex.ai/en/stable/examples/embeddings/OpenAI/}, 2024,
  accessed: 2025-07-25.

\bibitem{65}
{OpenAI}, ``New embedding models and api updates,''
  \url{https://openai.com/blog/new-embedding-models-and-api-updates}, 2024,
  accessed: 2025-07-25.

\bibitem{hasan2025llm}
S.~Hasan and A.~Rezai, ``Llm: Retreival vs. parametricmemory tradeoff: A
  comparison of retrieval-augmented generation and standalone largelanguage
  models using ragas answer accuracy,'' 2025.

\bibitem{kojima2022large}
T.~Kojima, S.~S. Gu, M.~Reid, Y.~Matsuo, and Y.~Iwasawa, ``Large language
  models are zero-shot reasoners,'' \emph{Advances in neural information
  processing systems}, vol.~35, pp. 22\,199--22\,213, 2022.

\bibitem{khot2022decomposed}
T.~Khot, H.~Trivedi, M.~Finlayson, Y.~Fu, K.~Richardson, P.~Clark, and
  A.~Sabharwal, ``Decomposed prompting: A modular approach for solving complex
  tasks,'' \emph{arXiv preprint arXiv:2210.02406}, 2022.

\bibitem{vatsal2024survey}
S.~Vatsal and H.~Dubey, ``A survey of prompt engineering methods in large
  language models for different nlp tasks,'' \emph{arXiv preprint
  arXiv:2407.12994}, 2024.

\bibitem{dua2022successive}
D.~Dua, S.~Gupta, S.~Singh, and M.~Gardner, ``Successive prompting for
  decomposing complex questions,'' \emph{arXiv preprint arXiv:2212.04092},
  2022.

\bibitem{press2022measuring}
O.~Press, M.~Zhang, S.~Min, L.~Schmidt, N.~A. Smith, and M.~Lewis, ``Measuring
  and narrowing the compositionality gap in language models,'' \emph{arXiv
  preprint arXiv:2210.03350}, 2022.

\bibitem{27}
L.~Wang, W.~Xu, Y.~Lan, Z.~Hu, Y.~Lan, R.~K.-W. Lee, and E.-P. Lim,
  ``Plan-and-solve prompting: Improving zero-shot chain-of-thought reasoning by
  large language models,'' \emph{arXiv preprint arXiv:2305.04091}, 2023.

\bibitem{69}
S.~Zheng, J.~Wang, Y.~Bai, S.~Wu, Y.~Du, X.~Li, C.~Xu, Y.~Zhang, J.~Ma, J.~Lin
  \emph{et~al.}, ``Judging llm-as-a-judge with mt-bench and chatbot arena,''
  \emph{arXiv preprint arXiv:2306.05685}, 2023.

\bibitem{70}
X.~Geng, A.~Chen, E.~Zhang, T.~Hashimoto, D.~Jurafsky, P.~Liang, T.~Zhang
  \emph{et~al.}, ``Koala: A dialogue model for academic research,'' \emph{arXiv
  preprint arXiv:2304.14108}, 2023.

\bibitem{abbasloo2025measuring}
S.~Abbasloo, ``Measuring reasoning in llms: a new dialectical angle,''
  \emph{arXiv preprint arXiv:2510.18134}, 2025.

\bibitem{10}
J.~Achiam, S.~Adler, S.~Agarwal, L.~Ahmad, I.~Akkaya, F.~L. Aleman, D.~Almeida,
  J.~Altenschmidt, S.~Altman, S.~Anadkat \emph{et~al.}, ``Gpt-4 technical
  report,'' \emph{arXiv preprint arXiv:2303.08774}, 2023.

\bibitem{srivastava2025towards}
G.~Srivastava, S.~Cao, and X.~Wang, ``Towards reasoning ability of small
  language models,'' \emph{arXiv preprint arXiv:2502.11569}, 2025.

\bibitem{thelwall2026can}
M.~Thelwall and E.~Mohammadi, ``Can small and reasoning large language models
  score journal articles for research quality and do averaging and few-shot
  help?'' \emph{Scientometrics}, pp. 1--33, 2026.

\bibitem{13}
\BIBentryALTinterwordspacing
{MITRE}, ``2024 cwe top 25 most dangerous software weaknesses,'' 2024,
  accessed: 2025-04-24. [Online]. Available:
  \url{https://cwe.mitre.org/top25/archive/2024/2024_cwe_top25.html}
\BIBentrySTDinterwordspacing

\bibitem{alansari2026large}
A.~Alansari and H.~Luqman, ``Large language models hallucination: A
  comprehensive survey,'' \emph{Computer Science Review}, vol.~61, p. 100970,
  2026.

\bibitem{song2026large}
P.~Song, P.~Han, and N.~Goodman, ``Large language model reasoning failures,''
  \emph{arXiv preprint arXiv:2602.06176}, 2026.

\bibitem{shen2025faithcot}
X.~Shen, S.~Wang, Z.~Tan, L.~Yao, X.~Zhao, K.~Xu, X.~Wang, and T.~Chen,
  ``Faithcot-bench: Benchmarking instance-level faithfulness of
  chain-of-thought reasoning,'' \emph{arXiv preprint arXiv:2510.04040}, 2025.

\bibitem{kaplan2020scaling}
J.~Kaplan, S.~McCandlish, T.~Henighan, T.~B. Brown, B.~Chess, R.~Child,
  S.~Gray, A.~Radford, J.~Wu, and D.~Amodei, ``Scaling laws for neural language
  models,'' \emph{arXiv preprint arXiv:2001.08361}, 2020.

\bibitem{wiki_gpt4_2024}
\BIBentryALTinterwordspacing
{Wikipedia contributors}, ``Gpt-4 --- {Wikipedia}{,} the free encyclopedia,''
  \url{https://en.wikipedia.org/wiki/GPT-4}, 2026, accessed: 2026-04-29.
  [Online]. Available: \url{https://en.wikipedia.org/wiki/GPT-4}
\BIBentrySTDinterwordspacing

\bibitem{hui2024qwen2}
B.~Hui, J.~Yang, Z.~Cui, J.~Yang, D.~Liu, L.~Zhang, T.~Liu, J.~Zhang, B.~Yu,
  K.~Lu \emph{et~al.}, ``Qwen2. 5-coder technical report,'' \emph{arXiv
  preprint arXiv:2409.12186}, 2024.

\bibitem{li2025everything}
Y.~Li, X.~Li, H.~Wu, M.~Xu, Y.~Zhang, X.~Cheng, F.~Xu, and S.~Zhong,
  ``Everything you wanted to know about llm-based vulnerability detection but
  were afraid to ask,'' \emph{arXiv preprint arXiv:2504.13474}, 2025.

\bibitem{71}
H.~Zheng, L.~Shen, Y.~Luo, T.~Liu, J.~Shen, and D.~Tao, ``Decomposed prompt
  decision transformer for efficient unseen task generalization,''
  \emph{Advances in Neural Information Processing Systems}, vol.~37, pp.
  122\,984--123\,006, 2024.

\bibitem{72}
T.~Khot, H.~Trivedi, M.~Finlayson, Y.~Fu, K.~Richardson, P.~Clark, and
  A.~Sabharwal, ``Decomposed prompting: A modular approach for solving complex
  tasks,'' \emph{arXiv preprint arXiv:2210.02406}, 2022.

\end{thebibliography}

% \clearpage
% % \appendix
% \input sec/Appendix.tex

\end{document}